\documentclass[USenglish,oneside,twocolumn]{article}

\usepackage[utf8]{inputenc}
\usepackage[big]{dgruyter_NEW}

\DOI{}

\usepackage{amsmath}
\usepackage{amssymb}
\usepackage{amsthm}
\usepackage{amsfonts}
\usepackage{graphicx}
\usepackage{siunitx}
\usepackage{multirow}
\usepackage[toc,page]{appendix}
\PassOptionsToPackage{hyphens}{url}\usepackage{hyperref}

\theoremstyle{plain}
\newtheorem{theorem}{Theorem}
\newtheorem*{theorem*}{Theorem}
\newtheorem{lemma}{Lemma}
\newtheorem*{lemma*}{Lemma}
\newtheorem*{corollary}{Corollary}

\newtheorem{definition}{Definition}
\newtheorem*{definition*}{Definition}

\begin{document}

%\setlength\topsep{0cm}
%\setlength\itemsep{0cm}
%\setlength\parsep{0cm}

%\author*[1]{Anonym}
\author*[1]{Bal\'azs Pej\'o}
\author[2]{Qiang Tang}
\author[3]{Gergely Bicz\'ok}

%\affil[1]{Anonym Institute}
\affil[1]{University of Luxembourg, E-mail: balazs.pejo@uni.lu}
\affil[2]{Luxembourg Institute of Science and Technology, E-mail: qiang.tang@list.lu}
\affil[3]{CrySyS Lab, Dept. of Networked Systems and Services, Budapest Univ. of Technology and Economics, E-mail: biczok@crysys.hu}

\title{\huge Together or Alone: The Price of Privacy in Collaborative Learning}
\runningtitle{Together or Alone: The Price of Privacy in Collaborative Learning}

	\begin{abstract}
		{Machine learning algorithms have reached mainstream status and are widely deployed in many applications. The accuracy of such algorithms depends significantly on the size of the underlying training dataset; in reality a small or medium sized organization often does not have the necessary data to train a reasonably accurate model. For such organizations, a realistic solution is to train their machine learning models based on their joint dataset (which is a union of the individual ones). Unfortunately, privacy concerns prevent them from straightforwardly doing so. While a number of privacy-preserving solutions exist for collaborating organizations to securely aggregate the parameters in the process of training the models, we are not aware of any work that provides a rational framework for the participants to precisely balance the privacy loss and accuracy gain in their collaboration.\\
        In this paper, by focusing on a two-player setting, we model the collaborative training process as a two-player game where each player aims to achieve higher accuracy while preserving the privacy of its own dataset. We introduce the notion of \textit{Price of Privacy}, a novel approach for measuring the impact of privacy protection on the accuracy in the proposed framework. Furthermore, we develop a game-theoretical model for different player types, and then either find or prove the existence of a Nash Equilibrium with regard to the strength of privacy protection for each player. Using recommendation systems as our main use case, we demonstrate how two players can make practical use of the proposed theoretical framework, including setting up the parameters and approximating the non-trivial Nash Equilibrium.}
	\end{abstract}

    \vspace{-0.5cm}

	%\journalname{Proceedings on Privacy Enhancing Technologies}
	
    \DOI{}
    \startpage{1}
    \received{}
    \revised{}
    \accepted{}

    \journalyear{}
    \journalvolume{}
    \journalissue{}

	\maketitle

    \vspace{-0.5cm}
    \section{Introduction}
    \label{sec:int}
    \vspace{-0.25cm}
    
\begin{figure}[h]
        \centering
        \includegraphics[width=3.3cm]{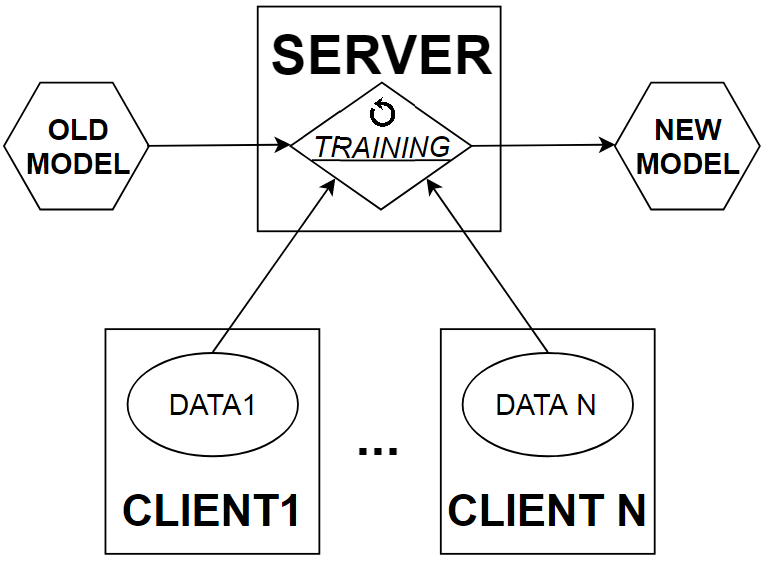}
        \includegraphics[width=4.4cm]{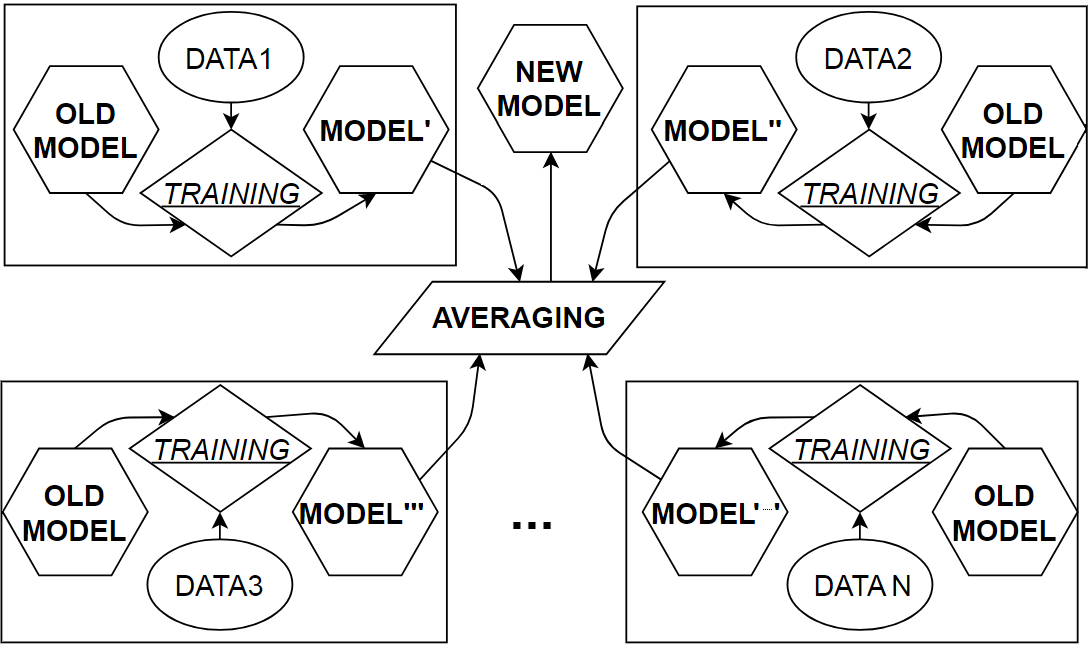}
        \caption{Centralized (left) and Distributed (right) Learning}
        \label{fig:ML}
    \end{figure}

    As data have become more valuable than oil, everybody wants to have a slice of it; Internet giants (e.g., Amazon, Google, Netflix, etc.) and small businesses alike would like to extract as much value from it as possible. Machine Learning (the process of learning from data and making predictions about it by building a model) has received much attention over the last decade, mostly due to its vast application range such as recommendation services, medicine, speech recognition, banking, gaming, driving, and more. For Machine Learning tasks, it is widely known that more training data will lead to a more accurate model. Unfortunately, most organizations do not possess a dataset as large as Netflix's or Amazon's. In such a situation, to obtain a relatively accurate model, a natural solution would be to aggregate all the data from different organizations on a centralized server and train on the global dataset as seen on the left side of Fig. \ref{fig:ML}. This approach is efficient, however, data owners have a valid privacy concern about sharing their data, particularly with new privacy regulations such as the European General Data Protection Regulation (GDPR). Therefore,  improving Machine Learning via straightforward data aggregation is likely undesirable and potentially unlawful in reality. Various privacy concerns exists with regard to Machine Learning (e.g., the privacy of the input to the training or the privacy of the trained model); in this paper, we focus on the privacy of the input for individual data contributors.
    
    \vspace{-0.25cm}
    \vspace{-0.25cm}
    
    In the literature, Privacy Preserving Distributed Machine Learning \cite{pathak2010multiparty,rajkumar2012differentially,hamm2016learning,mcmahan2016communication,pawlick2016stackelberg} have been proposed to solve this problem by training the model locally and safely aggregating all the local updates, illustrated on the right side of Fig. \ref{fig:ML}. On the other hand, these approaches' efficiency depend on the number of participants and the sample sizes as we highlight this in the related works.
    
    %For example, in Federated Learning \cite{mcmahan2016communication} the players add pairwise noise to their model before aggregation. By doing so, they successfully hide their model from each other. Also, the pairwise noises cancel each other out during aggregation such that the final model is not noisy. On the other hand, if there is a very small number of participants (e.g., two), this method does not provide privacy since the participants know exactly the noise added by the other.
    
    %Consequently, in this paper, we are interested in a scenario where only two participants are present, and they hold a significant part of the overall data.
    
    %PP-DML has some appealing properties like parallelization and privacy preservation, although it does not consider the participants with hostile samples. To handle the malicious participation, adversarial learning (AL) \cite{charikar2016learning} and byzantine failure tolerant learning (BFT-L) \cite{blanchard2017byzantine,chen2017distributed} were introduced recently.
    
    %Due to the safe aggregation in PP-DML, there is no way to value the contribution of the data owners individually. Consequently, freeriding -- participating in the distributed training without any real contribution -- is an issue. For example, instead of training the model using his/her local dataset (s)he just echoes back the old model - is an issue. Even in AL/BFT-L, the malicious inputs are only mitigated, not identified (and possibly punished).
    
    In this paper, we are interested in a scenario with two participants, each of whom possesses a significant amount of data and would like to obtain a more accurate model than what they would obtain if training was carried out in isolation. It is clear that the players will only be interested in collaboration if they can actually benefit from each other. To this end, we simply assume that the players have already evaluated the quality of each other's datasets to make sure training together is beneficial for both of them before the collaboration. How such evaluation should be done is out of scope for our research; there are best practices already established in the field \cite{dmbook}. Most of the Machine Learning papers, including privacy-preserving ones, implicitly make this assumption. %In addition, addressing the issues in scenarios with more participants is the future work for the next step. 
    
    %	\begin{example}
    %		Assume Netflix\footnote{\url{https://www.netflix.com}} and Amazon\footnote{\url{https://www.primevideo.com}} would like to increase their recommendation accuracy by collaboratively training a model with each other. They have different goals: instead of aiming for a model representing the whole population, they would like to increase the accuracy for their local users. Such collaboration raises privacy concerns, so to prevent the information leakage of their datasets, they must employ some privacy protection mechanism based on what 'privacy' means for them.
    %	\end{example}
    
    \vspace{-0.5cm}
    \subsection{Problem Statement}
    \vspace{-0.25cm}
    
    %How they determine this is not the primary focus of the paper, although we present an exemplary method in Sec. \ref{sec:approx} to tackle this problem. 
    
    Collaborative Machine Learning will increase the model accuracy, but at the cost of leaking some information about the players' datasets to each other. To mitigate the information leakage, players can apply some privacy-preserving mechanisms, e.g., calibrating and adding some noise or deleting some sensitive attributes. Many ``solutions'' have been proposed, as surveyed in the related work. In most of them, the players are not provided with the option of choosing their own privacy parameters. Clearly, there is a gap between these solutions and reality, where players will have different preferences to privacy and utility and may want to dynamically set the parameters.
    
    To bridge this gap, we consider the parties involved as rational players and model their collaboration as a two-player game. In our setting, players have their own trade-offs with respect to their privacy and expected utility and can flexibly set their own privacy parameters. The central research problem is to propose a general game theoretical model and find a Nash Equilibrium. Moreover, given a specific Machine Learning task, we should answer the following core questions. 
    
    \begin{itemize}
        \item What are the potential ranges for privacy parameters that make the collaborative Machine Learning model more accurate than training alone?
        \item What is the optimal privacy parameter (which results in the highest payoff)?
        \item With this optimal parameter, how much accuracy is lost overall due to the applied privacy-preserving mechanisms?
    \end{itemize}
    \vspace{-0.0cm}
    
    \vspace{-0.5cm}
    \subsection{Contribution}
    \vspace{-0.25cm}
    
    We first propose a two-player game theoretical model for \textit{Collaborative Learning} (a training process via an arbitrary training algorithm between two players). We profile the players and analyze their best response strategies and the equilibria of the designed game. Inspired by the notion of Price of Anarchy \cite{koutsoupias1999worst}, we define \textit{Price of Privacy}, which is a new way of measuring the accuracy degradation due to privacy protection. Then, we demonstrate the usage of the model via a recommender use case, where two players improve their own recommendation accuracy by leveraging on each other's dataset. It is worth noting that this is indeed a representative example since the used Stochastic Gradient Descent optimization process is a universal procedure widely used in Machine Learning tasks. For illustration purposes, we consider two privacy preserving mechanisms, including attribute deletion and differential privacy. Based on heuristics, we demonstrate how to approximate the privacy-accuracy trade-off functions, which lie in the core of the proposed theoretical model and determine how the players should set the parameters, and illustrate the practically obtained Nash Equilibrium. 
    
    We would like to emphasize that approximating the privacy-accuracy trade-off function is a very realistic choice in applying the proposed theoretical model. Scientifically, we may want to use cryptographic techniques such as secure two-party computation protocols to precisely compute these parameters. However, this is undesirable due to the incurred complexity. In order to reduce complexity, most deployed Machine Learning systems implement heuristics, such as approximating the parameters in Stochastic Gradient Descent \cite{dmbook}. 
    
    %This way the contribution of the players are measurable, while the privacy leakage can be mitigated via manipulating the data\footnote{More detail in Sec. \ref{sec:man}}. Table \ref{tab:comp} summarizes the learning methods and their advantages/disadvantages and shows the gap we would like to fill.
    
    %	\begin{table}[h]
    %		\centering
    %		\begin{tabular}{|c||c|c|c|}
    %			\hline
    %			& CL & ? & PP-DML\\
    %			\hline
    %			\hline
    %			Privacy & No & Some & Yes \\
    %			\hline
    %			Contribution & Known & Guessable & Unknown\\
    %			\hline
    %		\end{tabular}
    %		\caption{Learning with both privacy and measurable contribution}
    %		\label{tab:comp}
    %	\end{table}
    
    \vspace{-0.5cm}
    \subsection{Organization}
    \vspace{-0.25cm}
    
    In Sec. \ref{sec:intro_game}, we review some basic concepts used throughout the paper such as Game Theory and Differential Privacy. 
    In Sec. \ref{sec:game}, we introduce the Collaborative Learning game, explain the parameters, and define the concept of \textit{Price of Privacy}. 
    In Sec. \ref{sec:anal}, we provide a theoretical analysis of the proposed game and investigate the Nash Equilibrium. 
    In Sec. \ref{sec:set}, we introduce the recommender use case and describe two example privacy-preserving mechanisms. 
    In Sec. \ref{sec:exp}, for the recommender use case, we demonstrate how to determine the the privacy accuracy trade-off function via interpolation over the joint dataset. 
    Then, in Sec. \ref{sec:phi} we show the corresponding equilibrium by applying our game theoretic model. 
    In Sec. \ref{sec:approx}, besides presenting the whole process required in advance of the collaboration, we show how to approximate the trade-off function via heuristics and study its impact on the Nash Equilibrium. 
    In Sec. \ref{sec:related}, we review the the related works from the perspective of game theory and privacy-preserving machine learning. 
    In Sec. \ref{sec:conc}, we conclude the paper.
    
    As we use multiple well-known concepts  through the paper, we provide a short summary of abbreviations in App. \ref{app:abr} to improve readability. 
	
    \vspace{-0.5cm}
    \section{Preliminaries}
    \label{sec:intro_game}
    \vspace{-0.25cm}
    
    In this section, we introduce differential privacy and the game theoretic terminology used in the paper.
    
    \vspace{-0.5cm}
    \subsection{Differential Privacy}
    \vspace{-0.25cm}
    
    DP \cite{dwork2006calibrating} have been used widely in the literature. It classically quantifies the privacy of a mechanism in terms of parameters $\varepsilon$:
    
    \vspace{-0.25cm}
    \begin{definition*}[$\varepsilon$-differential privacy \cite{dwork2006calibrating}]
        An algorithm $\mathcal{A}$ is $\varepsilon$-DP ($\varepsilon\in[0,\infty)]$) if for any two datasets $D_1$ and $D_2$ that differ on a single element and for any set of possible outputs $O$:
        \vspace{-0.25cm}
        \begin{equation*}
        \Pr(\mathcal{A}(D_1)\in O)\leq e^{\varepsilon}\cdot\Pr(\mathcal{A}(D_2)\in O)
        \end{equation*}
    \end{definition*}
    \vspace{-0.25cm}
    
    DP gives a strong guarantee that presence or absence of a single data point will not change the final output of the algorithm significantly. Furthermore, the combination of DP mechanisms also satisfies DP:
    
    \vspace{-0.25cm}
    \begin{theorem*}[Composition Theorem \cite{dwork2006calibrating}]
        If the mechanisms $\mathcal{A}_i$ are $\varepsilon_i$-DP, then any sequential combination of them is $\sum_i\varepsilon_i$-DP.
    \end{theorem*}
    \vspace{-0.25cm}
    
    To achieve DP, noise must be added to the output of the algorithm. In most cases, this noise is drawn from a Laplacian distribution and it is proportional to the sensitivity of the algorithm itself:
    
    \vspace{-0.25cm}
    \begin{theorem*}[Laplace Mechanism \cite{dwork2006calibrating}]
        For $f:\mathcal{D}\rightarrow\mathbb{R}^k$, if $s$ is the sensitivity of $f$ (i.e., $s=\max_{D_1,D_2}||f(D_1)-f(D_2)||$ for any two datasets $D_1$ and $D_2$ that differ on a single element) then the mechanism $\mathcal{A}(D)=f(D)+Lap(\frac{s}{\varepsilon})$ with independently generated noise to each of the $k$ outputs enjoys $\varepsilon$-DP.
    \end{theorem*}
    \vspace{-0.25cm}
    
    \vspace{-0.5cm}
    \subsection{Game Theory}
    \vspace{-0.25cm}
    
    GT \cite{harsanyi1988general} is ``the study of mathematical models of conflict between intelligent, rational decision-makers''. Almost every multi-party interaction can be modeled as a game. In our case, these decision makers are the participants (players) of Collaborative Learning.
    
    \vspace{-0.25cm}
    \begin{definition*}[Game]
        A normal form representation of a game is a tuple $\langle\mathcal{N},\Sigma,\mathcal{U}\rangle$, where $\mathcal{N}=\{1,\dots,m\}$ is the set of players, $\Sigma=\{S_1,\dots,S_m\}$ where $S_i=\{s_1,s_2,\dots\}$ is the set of actions for player $i$ and $\mathcal{U}=\{u_1,\dots,u_m\}$ is the set of payoff functions.
    \end{definition*}
    \vspace{-0.25cm}
    
    A Best Response (BR) strategy gives the most favorable outcome for a player, taking other players' strategies as given:
    
    \vspace{-0.25cm}
    \begin{definition*}[Best Response]
        For a game $\langle\mathcal{N},\Sigma,\mathcal{U}\rangle$ the BR strategy for player $i$ for a given strategy vector $s_{-i}=(s_1,\dots,s_{i-1},s_{i+1},\dots,s_m)$ is $\hat{s}_i$ if $\forall s_{i_j}\in S_i$: $u_i(\hat{s}_i,s_{-i})\geq u_i(s_{i_j},s_{-i})$.
    \end{definition*}
    \vspace{-0.25cm}
    
    A Nash Equilibrium (NE) is a strategy vector where all the player's strategies are BR strategies. In other words, in a NE state every player makes the best/optimal decision for itself as long as the others' choices remain unchanged:
    
    \vspace{-0.25cm}
    \begin{definition*}[Nash Equilibrium]
        A pure-strategy NE of a game $\langle\mathcal{N},\Sigma,\mathcal{U}\rangle$ is a strategy vector $(s_1^*,\dots,s_m^*)$ where $s_i^*\in S_i$, such that for each player $i$ $\forall s_{i_j}\in S_i$: $u_i(s_i^*,s_{-i}^*)\geq u_i(s_{i_j},s_{-i}^*)$ where $s_{-i}^*=(s_1^*,\dots,s_{i-1}^*,s_{i+1}^*,\dots,s_m^*)$.
    \end{definition*}
    \vspace{-0.25cm}
    
    NE provides a way of predicting what will happen if several entities are making decisions at the same time where the outcome depends on the decisions of the others. The existence of a NE means no player will gain more by unilaterally changing its strategy at this unique state. 
    
    Another concept of GT is Social Optimum, which is a strategy vector that maximizes social welfare:
    
    \vspace{-0.25cm}
    \begin{definition*}[Social Optimum]
        The Social Optimum of a game $\langle\mathcal{N},\Sigma,\mathcal{U}\rangle$ is a strategy vector $(s'_1,\dots,s'_m)$ where $s'_i\in S_i$, such that
        \vspace{-0.1cm}
        \begin{equation*}
        \max_{s_1\in S_1,\dots,s_m\in S_m}\sum_{n\in\mathcal{N}}u_n(s_1,\dots,s_m)=\sum_{n\in\mathcal{N}}u_n(s'_1,\dots,s'_m)
        \end{equation*}
    \end{definition*}
    \vspace{-0.25cm}
    
    Despite the fact that no one can do better by changing strategy, NEs are not necessarily Social Optimums (as an example see Prisoner's Dilemma \cite{harsanyi1988general}). Price of Anarchy \cite{koutsoupias1999worst} measures the ratio between these two: how the efficiency of a system degrades due to the selfish behavior of its players:
    
    \vspace{-0.25cm}
    \begin{definition*}[Price of Anarchy]
        PoA of a game $\langle\mathcal{N},\Sigma,\mathcal{U}\rangle$ is
        \vspace{-0.1cm}
        \begin{equation*}
        PoA:=\frac{\max_{s\in S}\sum_{n\in \mathcal{N}}u_n(s)}{\min_{s^*\in S^*}\sum_{n\in \mathcal{N}}u_n(s^*)}
        \end{equation*}
        where $S=S_1\times\cdots\times S_m$ is the set of all possible outcomes while $S^*$ is the set of NEs.
    \end{definition*}
    \vspace{-0.25cm}

    \vspace{-0.5cm}
    \section{Game Theoretic Model}
    \label{sec:game}
    \vspace{-0.25cm}
    
    In this section, we describe the Collaborative Learning (CoL) game which captures the actions of two privacy-aware data holders in the scenario of applying an arbitrary privacy preserving mechanism and training algorithm on their datasets. We define the corresponding utility functions and elaborate on its components. Furthermore, we introduce the notion of \textit{Price of Privacy}, a novel measure of the effect of privacy protection on the accuracy of players.
    
    \vspace{-0.5cm}
    \subsection{The Collaborative Learning Game}
    \vspace{-0.25cm}
    
    At a high level, the players' goal in the CoL game is to maximize their utility, which is a function of the model accuracy and the privacy loss. We do not consider the adversarial aspect of players, hence the gain includes only the accuracy improvements on the model for a particular player as benefit (without the accuracy decrease of the other player\footnote{Extending the game for competing companies is an interesting future direction.}) while the cost is private information leakage: the trained model leaks some information about the local dataset used for training. 
    
    Players only choose the privacy parameters for a predetermined privacy preserving method $M$ (rather than choosing the method \textit{and} the parameter). This means each $M$ corresponds to a different game with a different definition of privacy, rather than having one game where the players' actions are deciding which mechanism to use and to what extent. This is a restricted scenario, nonetheless, even this scenario barely lends itself to purely analytical treatment; it is already not straightforward to derive the exact NE. 
    
    \vspace{-0.25cm}
    \begin{table}[h]
        \centering
        \begin{tabular}{|c|l|}
            \hline
            \bf{Variable} & \bf{Meaning}\\
            \hline
            $M$ & Privacy mechanism applied by the players\\
            \hline
            $p_n$ & Privacy parameter for player $n$\\
            \hline
            $C_n^M$ & Privacy weight for player $n$\\
            \hline
            $B_n$ & Accuracy weight for player $n$\\
            \hline
            $\theta_n$ & Accuracy by training alone for player $n$\\
            \hline
            $\Phi_n^M(p_1,p_2)$ & Accuracy by training together for player $n$\\
            \hline
            $b(\theta_n,\Phi_n^M)$ & Benefit function for player $n$\\
            \hline
            $c^M(p_n)$ & Privacy loss function for player $n$\\
            \hline
        \end{tabular}
        \vspace{0.1cm}
        \caption{Parameters of the CoL game}
        \label{tab:param}
    \end{table}
    \vspace{-0.75cm}
    
    The variables of the CoL game are listed in Tab. \ref{tab:param}, where the accuracy is measured as the prediction error of the trained model: lower $\theta_n$ and $\Phi_n^M$ correspond to a more accurate model. Maximal privacy protection is represented via $p_n=1$, while $p_n=0$ means no protection for player $n$. The benefit and the privacy loss are not on the same scale as the first depends on the accuracy while the latter on information loss. To make them comparable, we introduce weight parameters: the benefit function is multiplied with the accuracy weight $B_n>0$, while the privacy loss function is multiplied with the privacy weight $C_n^M\geq0$.
    
    The collaborative accuracy $\Phi^M_n(p_1,p_2)$ naturally depends also on the datasets and the used algorithm besides the privacy parameters $p_n$ and the corresponding privacy mechanism $M$. However, for simplicity we abstract them since it does not affect our theoretical analysis as long as $\Phi^M_n$ is symbolic. %Note, that the we treat accuracy as an individual's property rather than a public good as in \cite{ioannidis2013linear,chessa2015game}. Another difference is the scope of the model: it can be applied to any learning algorithm rather than specific tasks such as scalar averaging.
    
    \vspace{-0.25cm}
    \begin{definition}[Collaborative Learning game]
        The CoL game is a tuple $\langle\mathcal{N},\Sigma,\mathcal{U}\rangle$, where the set of players is $\mathcal{N}=\{1,2\}$, their actions are $\Sigma=\{p_1,p_2\}$ where $p_1,p_2\in[0,1]$ while their utility functions are $\mathcal{U}=\{u_1,u_2\}$ such that for $n\in\mathcal{N}$:
        \vspace{-0.1cm}
        \begin{equation}
        \label{eq:ut}
        u_n(p_1,p_2)=B_n\cdot b(\theta_n,\Phi_n^M(p_1,p_2))-C_n^M\cdot c^M(p_n)
        \end{equation}
    \end{definition}
    \vspace{-0.25cm}
    
    The CoL game is of symmetric information, i.e.,  the introduced parameters are public knowledge (i.e., $M$, $B_n$, $C_n^M$, $b$, $c^M$, $\theta_n$ and $\Phi_n^M$) except for the actions of the players (i.e., $p_n$). Moreover, we do not consider any negative effect of the training such as time or electricity consumption, however, such variables may be introduced to the model in the future. 
    
    In the following, whenever possible, we simplify the notion $C_n^M$, $c^M$ and $\Phi_n^M$ by removing the symbol $M$ to use $C_n$, $c$ and $\Phi_n$ respectively. We only need to keep in mind that these functions depend on the underlying privacy-preserving mechanism $M$ in the implementation. 
    
    \vspace{-0.5cm}
    \subsubsection{Privacy Loss Function $c^M(p_n)$}
    \label{sec:pri}
    \vspace{-0.25cm}
    
    This function represents the loss due to private data leakage. We define $c$ with the following natural properties:
    
    \vspace{-0.25cm}
    \begin{definition}[Privacy loss function]
        \label{def:c}
        $c:[0,1]\rightarrow[0,1]$ such that it is continuous and twice differentiable, $c(0)=1$, $c(1)=0$ and $\partial_{p_n}c<0$.
    \end{definition}
    \vspace{-0.25cm}
    
    This definition indicates that the maximal potential leakage is $1$ which corresponds to no protection at all, while maximal privacy protection corresponds to zero privacy loss. Furthermore, $c$ is monotone decreasing which means more privacy protection corresponds to less privacy loss.
    
    \vspace{-0.5cm}
    \subsubsection{Benefit Function $b(\theta_n,\Phi_n^M)$}
    \vspace{-0.25cm}
    
    The benefit function has two inputs: the accuracy achieved by training alone ($\theta_n$) and when both players train together ($\Phi_n$). Since a rational player would not collaborate to end up with a model of lower accuracy, we are only interested in the case when $\Phi_n<\theta_n$ (Note, that accuracy is measured as prediction error: lower values correspond to higher accuracy.). We define $b$ with the following natural properties:
    
    \vspace{-0.25cm}
    \begin{definition}[Benefit function]
        \label{def:b}
        $b:\mathbb{R}^+\times\mathbb{R}^+\rightarrow\mathbb{R}_0^+$ such that it is continuous and twice differentiable, $\partial_{p_n}b\leq0$ and $b(\theta_n, \Phi_n)=0$ if $\theta_n\leq\Phi_n$.
    \end{definition}
    \vspace{-0.25cm}
    
    This definition indicates that there is no benefit when the accuracy of the model trained together is lower than the accuracy of the model trained alone. Furthermore, since $b$ is monotone decreasing in $p_n$, more privacy protection results in lower benefit due to decreased accuracy.
    
    \vspace{-0.5cm}
    \subsubsection{Privacy-Accuracy Trade-off Function $\Phi_n^M(p_1,p_2)$}
    \label{sec:acc}
    \vspace{-0.25cm}
    
    $\Phi_n$ plays a crucial role in the benefit function $b$. However, the function of how a privacy protection mechanism affects a complex training algorithm (and consequently the accuracy) is unique for each dataset and algorithm. Although we measure it in Sec. \ref{sec:exp}, interpolate it in Sec. \ref{sec:phi} and approximate it in Sec. \ref{sec:approx} for a recommendation system use case, in general the exact form of $\Phi_n$ is unknown. On the other hand, some properties are expected:
    
    \vspace{-0.25cm}
    \begin{definition}[Privacy-Accuracy trade-off function]
        \label{def:phi}
        $\Phi_n:[0,1]\times[0,1]\rightarrow\mathbb{R}^+$ such that it is continuous, twice differentiable and:
        \begin{itemize}
            \item $\exists m\in\mathcal{N}:p_m=1\Rightarrow\forall n\in\mathcal{N}:\Phi_n(p_1,p_2)\geq\theta_n$
            \item $\forall n,m\in\mathcal{N}: \partial_{p_m}\Phi_n>0$
            \item $\forall n\in\mathcal{N}: \theta_n>\Phi_n(0,0)$
        \end{itemize}
    \end{definition}
    \vspace{-0.25cm}
    
    The first property means that maximal privacy protection cannot result in higher accuracy than training alone for any player. The second property indicates that higher privacy protection corresponds to lower accuracy since $\Phi_n$ is monotone increasing in both $p_1$ and $p_2$. The last property ensures that training together with no privacy corresponds to higher accuracy than training alone.
    
    \vspace{-0.5cm}
    \subsection{The Concept of Price of Privacy}
    \vspace{-0.25cm}
    
    Inspired by the notion of Price of Anarchy \cite{koutsoupias1999worst}, we define \textit{Price of Privacy} to measure the accuracy loss due to privacy constraints:
    
    \vspace{-0.25cm}
    \begin{definition}[Price of Privacy]
        \label{def:pop}
        $PoP$ measures the overall effect of privacy protection on the accuracy: 
        \vspace{-0.1cm}
        \begin{equation}
        \label{eq:pop}	PoP(p_1^*,p_2^*):=1-\frac{\sum_nb(\theta_n,\Phi_n(p_1^*,p_2^*))}{\sum_nb(\theta_n,\Phi_n(0,0))}
        \end{equation}
        The quotient is between the total accuracy improvement in a NE $(p_1^*,p_2^*)$ and the total accuracy improvement without privacy protection.
    \end{definition}
    \vspace{-0.25cm}
    
    Due to the Def. \ref{def:b} and \ref{def:phi}, $PoP\in[0,1]$ where 0 corresponds the highest possible accuracy which can be achieved via collaboration with no privacy while 1 corresponds the lowest possible accuracy which can be achieved by training alone. In other words, \textit{Price of Privacy} evaluates the benefit of a given equilibrium. The lower its value is, the higher the accuracy achieved by collaboration. %$PoP=0$ means $0\%$ degradation in accuracy due to privacy protection (no price paid in accuracy) while $PoP=1$ corresponds to $100\%$ degradation meaning all the possible accuracy improvement is lost due to the applied privacy mechanism.
    
    Note that while PoA characterizes the whole game, $PoP$ is a property of a NE. Also, since $\Phi_n$ can only be estimated in a real-world scenario, the players can only approximate the value of $PoP$, which would then measure the efficiency of the collaboration.
    
    \vspace{-0.5cm}
    \subsection{Remarks on the Model}
    \vspace{-0.25cm}
    
    Given that the actual value of $\Phi_n$ is required to compute the optimal strategies, $\Phi_n$ has to be numerically evaluated for putting the CoL game to practical use.
    Different from other parameters which can be set freely, the impact of the privacy-preserving mechanism $M$ on the joint accuracy (and thus $\Phi_n$) is determined by both datasets. Precisely computing this function requires access to the joint dataset; thus, it raises the very privacy concern which we want to mitigate in the first place. To break this loop, we propose to adopt an approximation approach for applying the model. To this end, we provide a solution heuristic and show its practical feasibility in Sec. \ref{sec:approx}.  

    \vspace{-0.5cm}
    \section{Equilibrium Analysis}
    \label{sec:anal}
    \vspace{-0.25cm}
    
    %This section answers the questions what is the optimal privacy parameter ($Q_2$) and how much accuracy is lost due to privacy protection ($Q_3$)? In short, the answer is provided via NE for $Q_2$ while $PoP$ answers to $Q_3$. To improve readability all proofs are moved to Appendix \ref{app:proofs}.
    
    %Note, that a BR ($\hat{p}_n$) is not necessarily equal with the NE ($p_n^*$), as it can correspond to negative utility in which case the NE is no collaboration. 
    
    %\begin{remark}
    %	Note, that since our model of the training process is general, we cannot hope to provide closed form solutions -- even with predetermined benefit and privacy loss functions -- because that was not possible for simpler cases already such as linear regression \cite{ioannidis2013linear} or scalar averaging \cite{chessa2015game}.
    %\end{remark}
    
    In this section, we characterize the NEs for a simple and more complex cases of the CoL game. We derive symbolic NEs in closed form for the case where exactly one of the players is privacy-concerned (i.e., Collaboration-as-a-Service scenario). Next, we prove the existence of a pure strategy NE in the general case, where both players are privacy-concerned to a given degree. %We develop methods for estimating the numerical value of $\Phi_n$ and thus give practical numerical approximations of NEs for fully-specified CoL instance in Sec. \ref{sec:approx}. 
    To preserve clarity, all mathematical proofs for theorems in this section are given in App. \ref{app:proofs}.
    
    The simplest NE of the CoL game is no collaboration:
    
    \vspace{-0.25cm}
    \begin{theorem}
        \label{th:tri}
        Applying maximal privacy protection (training alone) in the CoL game is a NE: $(p^*_1,p^*_2)=(1,1)$.
    \end{theorem}
    \vspace{-0.25cm}
    
    Clearly, when the players train alone there will be no improvement in accuracy. This means that the \textit{Price of Privacy} for this NE is the maximum 1: the entire potential accuracy improvement is lost due to privacy protection. This finding seemingly contradicts  \cite{chessa2015game}, which states that all players refraining to participate cannot be an equilibrium. There is a significant difference though; estimation cost is a public good in \cite{chessa2015game}, while in our case accuracy is private and each participant has a base accuracy level obtained by training alone.
    
    \vspace{-0.5cm}
    \subsection{Player Types}
    \vspace{-0.25cm}
    
    Based on the properties of CoL game, two natural expectations arise:
    
    \begin{itemize}
        \item A player prefers collaboration if it values accuracy significantly more than privacy ($B_n\gg C_n$).
        \item A player prefers training alone if it values accuracy significantly less than privacy ($B_n\ll C_n$).
    \end{itemize}
    \vspace{-0.25cm}
    
    These intuitions are captured in the following two lemmas:
    
    \vspace{-0.25cm}
    \begin{lemma}
        \label{lemma:1}
        $\exists\alpha_n\geq0$ such that if $\frac{C_n}{B_n}\leq\alpha_n$ for player $n$ than its BR is $\hat{p}_n=0$.
    \end{lemma}
    \vspace{-0.75cm}
    \begin{lemma}
        \label{lemma:2}
        $\exists\beta_n\geq0$ such that if $\frac{C_n}{B_n}\geq\beta_n$ for player $n$ then its BR is to set $\hat{p}_n=1$.
    \end{lemma}
    \vspace{-0.25cm}
    
    The questions we are interested in answering are: \textit{what are the exact values of $\alpha_n$ and $\beta_n$} and \textit{what is the NE in case $\alpha_n\leq\frac{C_n}{B_n}\leq\beta_n$}. Based on the ratio $\frac{C_n}{B_n}\in[0,\infty]$, we define two types of players:
    
    \begin{itemize}
        \item \textbf{Unconcerned}: This type of player cares only about accuracy. This represents the case when $\frac{C_n}{B_n}=0$: the privacy weight for player $n$ is zero ($C_n=0$).
        \item \textbf{Concerned}: This player is more privacy-aware, as the privacy loss is present in its utility function. This represents the case when $\frac{C_n}{B_n}>0$.
    \end{itemize}
    \vspace{-0.25cm}
    
    This information is available to both players as the CoL game is a symmetric information game: both players know which type of player they face.
    
    \vspace{-0.5cm}
    \subsection{One Player is Privacy Concerned}
    \vspace{-0.25cm}
    \begin{definition}[Collaboration-as-a-Service]
        \label{def:caas}
        In a CaaS scenario one player acts as a for-profit service provider of collaborative training without privacy concerns, i.e., its privacy weight is 0. 
    \end{definition}
    \vspace{-0.25cm}
    
    \textbf{Example.} \textit{Imagine a company who offers CaaS for her own profit (Player 2). The CaaS provider does not apply any privacy-preserving mechanism (see Th. \ref{th:unc}). Any interested party (Player 1) who wants to to boost its accuracy can use this service. At the same time, Player 1 requires additional privacy protection (besides the inherent complexity of the training algorithm) to prevent her own data from leaking.}
    
    \vspace{-0.25cm}
    \begin{theorem}[Training as an unconcerned player]
        \label{th:unc}
        If player $n$ is unconcerned ($C_n=0$) then its BR is to collaborate without any privacy protection: $\hat{p}_n=0$.
    \end{theorem}
    \vspace{-0.25cm}
    
    When both players are unconcerned ($C_1=C_2=0$), $(p_1^*,p_2^*)=(0,0)$ is a NE. The corresponding \textit{Price of Privacy} value is 0 as no accuracy is lost due to privacy protection.
    
    As a result, the unconcerned player do not apply any privacy-preserving mechanism. Without loss of generality we assume Player 2 is unconcerned, so its BR is $\hat{p}_2=0$. This allows us to make the following simplifications: $\Phi(p_1):=\Phi_1(p_1,\hat{p}_2)$, $b(p_1):=b(\theta_1,\Phi(p_1,\hat{p}_2))$ and $u(p_1):=u_1(p_1,\hat{p}_2)$ while $f'=\partial_{p_1}f$ and $f''=\partial^2_{p_1}f$.
    
    \vspace{-0.25cm}
    \begin{theorem}[Training with an unconcerned player]
        \label{th:c_vs_uc_gen}
        A NE of the CoL game when Player 1 is concerned ($C_1>0$) while Player 2 is unconcerned ($C_2=0$) is $(p_1^*,p_2^*)=(\rho,0)$ where $\rho$ is defined by Eq. (\ref{eq:th_gen}) where $[\cdot]^{-1}$ is the inverse function of $[\cdot]$ and $r=\frac{C_1}{B_1}$:
        \vspace{-0.1cm}
        \begin{equation}
        \label{eq:th_gen}
        \rho=\left\{
        \begin{tabular}{ccc} \multirow{3}{*}{$\left[\frac{b'\Phi'}{c'}\right]^{-1}(r)$} & \multirow{3}{*}{if} & $u''(\rho)<0$\\
        & & $\rho\in[0,1]$\\
        & & $u(\rho)>0$\\
        $0$ & if & $b(0)>r$\\
        $1$ &  & otherwise \\
        \end{tabular}\right.
        \end{equation}
    \end{theorem}
    \vspace{-0.25cm}
    
    The three possible NEs when Player 2 is unconcerned, and the corresponding \textit{Price of Privacy} values are:
    
    \begin{itemize}
        \item If the possible maximal benefit is higher than the weight ratio ($b(0)>\frac{C_1}{B_1}$) for Player 1, this player should train without any privacy protection since $(p_1^*,p_2^*)=(0,0)$ is a NE. In this case $PoP=0$.
        \item If all the required conditions in Th. \ref{th:c_vs_uc_gen}  hold, $(p_1^*,p_2^*)=\left(\left[\frac{b'\Phi'}{c'}\right]^{-1}\left(\frac{C_1}{B_1}\right),0\right)$ is a NE with
        \vspace{-0.1cm}
        \begin{equation*}	
        PoP=1-\frac{\sum_nb\left(\theta_n,\Phi_n\left(\left[\frac{b'\Phi'}{c'}\right]^{-1}\left(\frac{C_1}{B_1}\right),0\right)\right)}{\sum_nb(\theta_n,\Phi_n(0,0))}
        \end{equation*}
        \item Otherwise $(p_1^*,p_2^*)=(1,0)$ is a NE. In this case, when one player apply maximal privacy protection (Player 1), the other player's utility cannot be positive due to the Def. \ref{def:phi}. Furthermore, since Player 2 is privacy unconcerned, its actual payoff is 0 independently of its action. As a result, $(p_1^*,p_2^*)=(1,p_2)$ is a NE for all $p_2\in[0,1]$ as they all correspond to the same 0 payoff. For simplicity, we use $(1,1)$ to represent this case (the players train alone), where the corresponding $PoP$ value is 1.
    \end{itemize}
    \vspace{-0.25cm}
    
    This result is quite abstract because all the components of the utility function are treated symbolically. However, even if we specify the benefit and the privacy loss functions, the privacy-accuracy trade-off function $\Phi_n$ would still be unknown due to the unspecified training algorithm. We show this in the next Cor. where we set $b$ and $c$ to be linear, as it is shown in Eq. (\ref{eq:spec}) where $[\cdot]^+=\max\{\cdot,0\}$.
    
    \vspace{-0.1cm}
    \begin{equation}
    \label{eq:spec}
    \begin{gathered}
    c(p_n):=1-p_n\\
    b(\theta_n,\Phi_n(p_1,p_2)):=[\theta_n-\Phi_n(p_1,p_2)]^+
    \end{gathered}
    \end{equation}
    
    \vspace{-0.25cm}
    \begin{corollary}
        With the same notations as in Th. \ref{th:c_vs_uc_gen} and with the benefit and privacy loss functions defined in Eq. (\ref{eq:spec}), $(p_1^*,p_2^*)=(\rho,0)$ is a NE when $C_1>0$ and $C_2=0$ if $\rho$ is:
        \vspace{-0.25cm}
        \begin{equation*}
        \rho=\hspace{-0.1cm}\left\{\hspace{-0.1cm}
        \begin{tabular}{ccc}
        $0$ & \hspace{-0.1cm}if\hspace{-0.1cm} & $r\leq\theta_1-\Phi(0)$\\
        $[\Phi']^{-1}\left(r\right)$ & \hspace{-0.1cm}if\hspace{-0.1cm} & $\Phi''\left(r\right)<0$, $u_1(\rho,0)>0$, $\rho\in[0,1]$\\
        $1$ & \hspace{-0.2cm} & otherwise \\
        \end{tabular}\right.
        \end{equation*}
    \end{corollary}
    \vspace{-0.25cm}
    
    To compute the exact (numerical) NE and the corresponding \textit{$PoP$}Price of Privacy of the CoL game, we need to define the function $\Phi_n$. While for simpler training algorithms $\Phi_n$ is known \cite{ioannidis2013linear,chessa2015game}, for more complex algorithms it can only be approximated. We demonstrate a potential approximation method called self-division in Sec. \ref{sec:approx}.
    
    \vspace{-0.5cm}
    \subsection{Both Players are Privacy Concerned}
    \label{sec:con_vc_con}
    \vspace{-0.25cm}
    
    Now we consider the general case when both players' privacy weights are non-zero. We prove the existence of a pure-strategy NE besides the trivial $(p_1^*,p_2^*)=(1,1)$; we utilize the chain rule of derivation for higher dimensions and a result from the theory of potential games \cite{monderer1996potential}.
    
    \vspace{-0.25cm}
    \begin{lemma*}[Chain Rule]
        If $f:\mathbb{R}^2\rightarrow\mathbb{R}$ and $g:\mathbb{R}^2\rightarrow\mathbb{R}$ are differentiable functions, then
        \vspace{-0.1cm}
        \begin{equation*}
        \forall i\in[1,2]:\partial_{x_i} f(x,g(x_1,x_2))=\partial_g f\cdot\partial_{x_i}g(x_1,x_2)
        \end{equation*}
    \end{lemma*}
    \vspace{-0.5cm}
    \begin{definition*}[Potential Game \cite{monderer1996potential}]
        A two-player game $G$ is a potential game if the mixed second order partial derivative of the utility functions are equal:
        \vspace{-0.1cm}
        \begin{equation}
        \label{def:pot}
        \partial_{p_1}\partial_{p_2}u_1=\partial_{p_1}\partial_{p_2}u_2
        \end{equation}
    \end{definition*}
    \vspace{-0.5cm}
    \begin{theorem*}[Monderer \& Shapley \cite{monderer1996potential}]
        Every potential game admits at least one pure-strategy NE.
    \end{theorem*}
    \vspace{-0.25cm}
    
    Now we can state the theorem which holds even if both players are privacy-concerned:
    
    \vspace{-0.25cm}
    \begin{theorem}
        \label{th:uc_vs_uc_gen}
        The CoL game has at least one non-trivial pure-strategy NE if
        \vspace{-0.25cm}
        \begin{equation}
        \label{eq:uc_vs_uc_gen}
        \partial^2_{\Phi}b\cdot(\partial_{p_1}\Phi_1-\partial_{p_2}\Phi_2)=\partial_{\Phi}b\cdot(\partial_{p_1}\partial_{p_2}\Phi_2-\partial_{p_1}\partial_{p_2}\Phi_1)
        \end{equation}
    \end{theorem}
    \vspace{-0.5cm}
    \begin{corollary}[\textbf{1}]
        If we assume $\partial^i_{p_1}\Phi_1=\partial^i_{p_2}\Phi_2$ for $i\in\{1,2\}$ then Th. \ref{th:uc_vs_uc_gen} holds.
    \end{corollary}
    \vspace{-0.25cm}
    
    The condition on the derivatives of $\Phi_n$ in Cor. 1 means that the player's accuracy changes the same way in relation to their own privacy parameter, independently from the other player's privacy parameter. In Sec. \ref{sec:exp} we measure the accuracy for multiple privacy parameter values and find that this is indeed the case. Moreover, we find that $\Phi_1\approx\Phi_2$ when the players have equal dataset sizes.\footnote{Note that these findings are empirical: based on the specific datasets/algorithm we used. For more details see Sec. \ref{sec:exp}.}
    
    \vspace{-0.5cm}
    \subsection{Remarks on the NEs}
    \vspace{-0.25cm}
    
    Note that a non-trivial pure-strategy NE does not necessarily correspond to positive payoffs: if $\exists n\in\mathcal{N}:u_n(p_1^*,p_2^*)<0$ then player $n$ would rather not collaborate. Instead, it would train alone and gain 0. This situation is plausible since 0 utility corresponds to the accuracy of the locally trained model. As such, an actual accuracy improvement is not trivial by collaboration, especially with additional privacy concerns (as we show in Sec. \ref{sec:exp}). 

    \vspace{-0.5cm}
    \section{Use Case: Recommendation System}
    \label{sec:set}
    \vspace{-0.25cm}
    
    In this section, we describe our recommender use case (RecSys), where Machine Learning (ML) is performed through Matrix Factorization (MF) via Stochastic Gradient Descent (SGD). Then, we introduce two example privacy-preserving mechanisms: Suppression (\textit{Sup}) and bounded DP (\textit{bDP}).
    
    \vspace{-0.5cm}
    \subsection{The Learning Process}
    \vspace{-0.5cm}
    \begin{figure}[h]
        \centering
        \includegraphics[width=8cm]{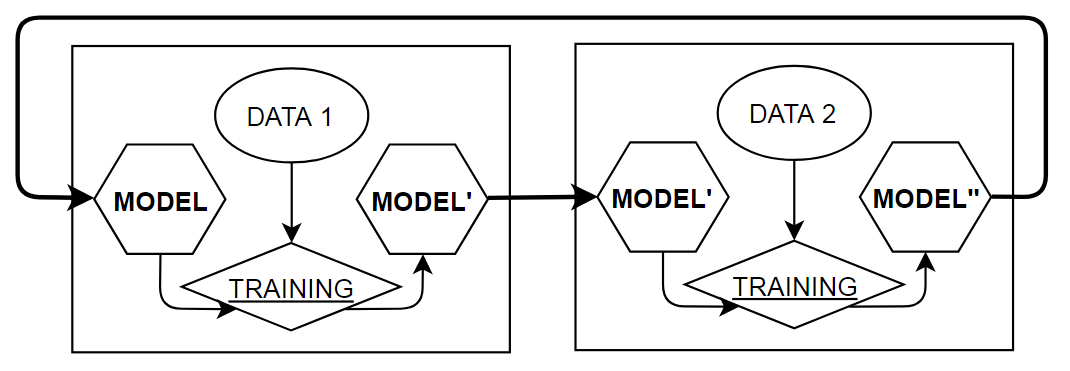}
        \caption{Learning Sequentially}
        \label{fig:itL}
    \end{figure}
    \vspace{-0.25cm}
    
    If there are only two parties in a distributed learning scenario (right side of Fig. \ref{fig:ML}), the trained model reveals some information about both players' dataset. Hence, in our scenario, the players train the same model iteratively without any safe aggregation. If there are only two participants, parallelization does not improve the efficiency much, so the players are training the model sequentially as seen in Fig. \ref{fig:itL}. The problem of information leakage is tackled with privacy-preserving mechanisms.
    
    Our use case is a RecSys scenario. We assume that players hold a user-item rating matrix with a common item-set $I=I_1=I_2$ and disjoint user-set: $U_1\cap U_2=\emptyset$ where $U=U_1\cup U_2$. As usual, $r_{ui}\in R_{|U|\times|I|}$ refers to the rating user $u$ gives item $i$.
    
\begin{figure}[h]
        \centering
        \includegraphics[width=8cm]{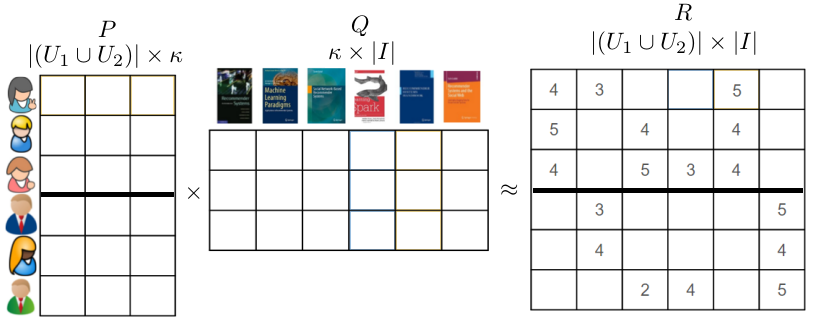}
        \caption{RecSyS Scenario}
        \label{fig:mf}
    \end{figure}

    The goal of the learning algorithm is to find the items that users desire. One of the most widespread method to do that is MF \cite{koren2009matrix} as seen in Fig. \ref{fig:mf}: finding $P_{|U|\times \kappa}$ and $Q_{\kappa\times|I|}$ such that $P\cdot Q\approx R$. As the user-sets are disjoint, players only need to share the item feature matrix $Q$. 
    
    \vspace{-0.25cm}
    \vspace{-0.25cm}
    
    The goal of a RecSys algorithm is to minimize the error between the prediction and the observed ratings as described in Eq. (\ref{eq:MF}) where $\lambda$ is the regularization parameter, while $p_u$ ($q_i$) is the corresponding row (column) in $P$ ($Q$) for $r_{ui}$.
    
    \vspace{-0.25cm}
    \begin{equation}
    \label{eq:MF}
    \min_{P,Q}\sum_{r_{ui}\in R}(r_{ui}-p_uq_i)^2+\lambda(||p_u||^2+||q_i||^2)
    \end{equation}
    
    One of the most popular techniques to minimize this formula is SGD. It works by iteratively selecting a random rating $r_{ui}\in R$ and updating the corresponding factor vectors according to Eq. (\ref{eq:upd}) where $e_{iu}=p_uq_i-r_{ui}$ and $\gamma$ is the learning rate.
    
    \vspace{-0.25cm}
    \begin{equation}
    \label{eq:upd}
    \begin{gathered}
    p'_{u}:=p_{u}+\gamma(e_{ui}q_{j}-\lambda p_{u})\\
    q'_{i}:=q_{i}+\gamma(e_{ui}p_{u}-\lambda q_{i})
    \end{gathered}
    \end{equation}
    
    Since we use SGD, the training process shown in Fig. \ref{fig:itL} is essentially equivalent to mini-batch learning where the batches are the datasets of the players. As we use RecSys as an illustrative example, we simplify it: we assume that players share the learning algorithm which is embedded with the necessary parameters such as learning rate $\gamma$, regularization parameter $\lambda$, number of features $\kappa$, and maximum number of iterations $\iota$. Notations are summarized in Tab. \ref{tab:sum}.
    
    \vspace{-0.25cm}
    \begin{table}[h]
        \centering
        \begin{tabular}{|c|l|}
            \hline
            \bf{Variable} & \bf{Meaning}\\
            \hline
            $U$ & Joint user-set\\
            \hline
            $I$ & Itemset\\
            \hline
            $R$ & Rating matrix\\
            \hline
            $r_{ui}$ & Rating of user $u$ for item $i$\\
            \hline
            $P,Q$ & Feature matrices\\
            \hline
            $\gamma$ & Learning rate\\
            \hline
            $\lambda$ & Regularization parameter\\
            \hline
            $\kappa$ & Number of features\\
            \hline
            $\iota$ & Number of iterations\\
            \hline
        \end{tabular}
        \vspace{0.1cm}
        \caption{RecSys Parameters}
        \label{tab:sum}
    \end{table}
    \vspace{-1cm}
    
    %\subsection{Deviation}
    
    %The players deviate from protocol based on the evaluation of the contribution of the other player. For example, the contribution can be the RMSE change when the other group is updating the model. We defined few parameters to be able to set different conditions when the players should deviate (i.e., stop participating in the Collaborative learning) from the protocol.
    
    %\begin{itemize}
    % \item \textbf{Buffer} ($b$): The deviating condition is checked on the last $b$ iteration. For example if $b=1$, only the last update's contribution is taken into account of the other player, while when $b=i_{max}$, the overall accumulated contribution is considered.
    % \item \textbf{Toleration} ($t$): Based on the last $b$ iteration, what is threshold of deviating from the collaboratively learning process.
    %\end{itemize}
    
    %Using these variables, we defined two conditions when the players should deviate from the protocol.
    
    %\begin{itemize}
    %	\item \textbf{Size}: Check the last $b$ iteration and check the accumulated effect of it. If it is smaller than $t$, stop cooperating. For example, $t=0$ means if the last $b$ iteration worsen the prediction, the player will no longer participate in the Collaborative learning process.
    %	\item \textbf{Direction}: Check the last $b$ iteration and count how many improved. If it is smaller than $t$, stop. For example, $t=0$ means if none of the last $b$ iterations improved, the player will no longer participate in the Collaborative learning process.
    %\end{itemize}
    
    \vspace{-0.5cm}
    \subsection{Privacy Preserving Mechanisms}
    \vspace{-0.25cm}
    
    We focus on input manipulation for  privacy preservation as we are concerned with input data privacy. In fact, \cite{friedman2016differential} concluded that input perturbation achieves the most efficient accuracy-privacy trade-off amongst various DP mechanisms. We investigate \textit{Sup} and \textit{bDP} as available mechanisms. %To mitigate the privacy loss, noise can be injected into the dataset in advance of the training. We considered two method:
    
    \vspace{-0.5cm}
    \subsubsection{Suppression}
    \vspace{-0.25cm}
    \begin{definition}[Suppression]
        Sup removes input data from the original dataset to protect it from information leakage resulting from the model or the learning process. The size of reduction is determined by the privacy parameter $p\in[0,1]$, i.e., $p$ is proportional to the data removed\footnote{We treat $p$ as a continuous variable even though it is discrete; this does not affect our analysis owing to large dataset sizes.}.
    \end{definition}
    \vspace{-0.25cm}
    
    \textit{Sup} essentially chooses a subset of the dataset to be used for training together. \textit{Sup} can be used to remove sensitive data from the dataset, so even if the other player can reconstruct the dataset from the trained model, the removed part remains fully protected.
    
    \vspace{-0.5cm}
    \subsubsection{Bounded Differential Privacy}
    \vspace{-0.25cm}
    
    To apply bDP, we must determine the sensitivity of the machine learning algorithm first:
    
    \vspace{-0.25cm}
    \begin{theorem}[Sensitivity of RecSys]
        \label{th:sens}
        The sensitivity $S$ of the introduced RecSys scenario is
        \vspace{-0.1cm}
        \begin{equation}
        \label{eq:sgd}
        S\leq\kappa\cdot\iota\cdot \gamma \cdot (\Delta r \cdot p_{max} -\lambda \cdot q_{max})
        \end{equation}
    \end{theorem}
    \vspace{-0.5cm}
    \begin{proof}[Proof Th. \ref{th:sens}]
        See App. \ref{app:proof2}.
    \end{proof}
    \vspace{-0.1cm}
    
    Now, we consider the \textit{bDP} mechanism \cite{friedman2016differential}:
    
    \vspace{-0.25cm}
    \begin{definition}[bounded DP]
        bDP aims to hide the value of a rating. To achieve $\varepsilon$-bDP, each rating is modified as it is shown in Eq. (\ref{eq:bdp}) where $L(x)$ is a Laplacian noise with 0 mean and $x$ variance.
        \vspace{-0.1cm}
        \begin{equation}
        \label{eq:bdp}
        r'_{ui}:=\hspace{-0.1cm}\left\{
        \begin{tabular}{lll}
        $r_{max}$ & if & $r_{ui}+L(\frac{S}{\varepsilon})\geq r_{max}$\\
        $r_{min}$ & if & $r_{ui}+L(\frac{S}{\varepsilon})\leq r_{min}$\\
        $r_{ui}+L(\frac{S}{\varepsilon})$ & & otherwise\\
        \end{tabular}\right.
        \end{equation}
    \end{definition}
    \vspace{-0.25cm}
    
    %		\begin{definition}[uDP]
    %			Unbounded DP aims to hide the existence of a rating. It works by inflating the original matrix with `fake' ratings and then applying \textit{bDP}. As the process is similar to \textit{bDP}, the exact method is only described in Appendix \ref{app:uDP}.
    %		\end{definition}
    
    \vspace{-0.5cm}
    \subsection{Unifying Privacy Parameters}
    \label{sec:pripar}
    \vspace{-0.25cm}
    
    These approaches are hard to compare since they focus on protecting different things. \textit{Sup} aims to provide maximal privacy for some of the data while leaving the rest unprotected. On the other hand, \textit{bDP} provides an equal amount of privacy for all the data based on the parameter $\varepsilon$. In the CoL game we defined the privacy parameter on a scale 0 to 1, therefore the specific parameters of \textit{Sup} and \textit{bDP} must be mapped to $[0,1]$ where $p=0$ means no privacy, while $p=1$ stands for full privacy protection. For \textit{Sup} the value $p$ is straightforward: $p$ represents the portion of data removed. Hiding the dataset in whole ($100\%$ protection) means $p=1$ while if the whole dataset is used for training ($0\%$ protection) then $p=0$.
    
    In case of \textit{bDP}, $100\%$ privacy ($p=1$) is achieved when $\varepsilon=0$ (infinite noise) while $\varepsilon\approx\infty$ corresponds to zero noise ($p=0$). This relation can be captured via a function $f:[0,\infty)\rightarrow[0,1]$ such that $f(0)=1$, $\lim_{x\rightarrow\infty}f(x)=0$ and $f$ is monotone decreasing. We use the mapping $p=f(\varepsilon)=\frac{1}{\varepsilon+1}$ and $\varepsilon=f^{-1}(p)=\frac1p-1$. This mapping does not carry meaning such as equivalence in any sense between methods, so it is not used for direct comparison. We only use it to convert the privacy parameter $\varepsilon$ into $[0,1]$ so we can use \textit{bDP} as privacy-preserving method $M$ in the CoL game defined in Sec. \ref{sec:game}.

    \vspace{-0.5cm}
    \section{Determining $\Phi$ for RecSys}
    \label{sec:exp}
    \vspace{-0.25cm}
    
    For all research questions in Sec. \ref{sec:int} answers depend on $\Phi_n^M$. Consequently, in this section we measure the model accuracy for various privacy parameters with regard to the learning task and the privacy mechanisms introduced in Sec. \ref{sec:set}. 
    
    For our experiments, we implemented SGD as training algorithm in Matlab \cite{git}. We used the MovieLens 1M \cite{movielens} and Netflix \cite{netflix} datasets; we shrunk the Netflix dataset to $10\%$ by randomly filtering out $90\%$ of the users. Furthermore, both datasets are preprocessed similarly to \cite{friedman2016differential}. Preprocessing is described in details in App. \ref{app:pre}. We will refer to the preprocessed datasets as 1M and NF10, respectively. The parameters of the preprocessed datasets are shown in Tab. \ref{tab:data}.
    
    \vspace{-0.25cm}
    \begin{table}[h]
        \centering
        \begin{tabular}{|c||c|c|c|c|}
            \hline
            Dataset & Rating & User & Item & Density\\
            \hline
            \hline
            %100k & \num{97953} & \num{943} & \num{1152} & \\
            %\hline
            1M & \num{998539} & \num{6040} & \num{3260} & $\num{0.051}$\\
            \hline
            %10m & \num{9995471} & \num{69878} & \num{9708} & \\
            %\hline
            NF10 & \num{10033823} & \num{46462} & \num{16395} & $\num{0.013}$\\
            \hline
        \end{tabular}
        \vspace{0.1cm}
        \caption{The Datasets Size after Preprocessing}
        \label{tab:data}
    \end{table}
    \vspace{-1cm}
    
    The algorithm for MF is SGD, where the number of features are $\kappa=4$. The algorithm runs for $20$ iterations ($\iota=20$) with learning rate $\gamma=\num{0.0075}$ and regularization parameter $\lambda=0.01$. The feature matrices are bounded by $p_{max}=q_{max}=0.5$. This means that the sensitivity of the RecSys scenario is $S\leq4\cdot20\cdot0.0075\cdot(2\cdot2\cdot0.5-0.01\cdot0.5)=1.197$ as a result of Th. \ref{th:sens}. %The initial feature matrices are following normal distribution.
    
    We assume if a model is trained using datasets from very different distributions, the model captures the properties of the mixed distribution of the combined dataset (which might be far from the original distributions). On the other hand, using training data from similar distributions results in capturing the statistical properties of a distribution close to the original ones. As such, if the players' datasets are from a similar different distribution, training together likely results in a more accurate model than training alone. Consequently, we imitate the players' datasets by splitting 1M and NF10: each user with its corresponding ratings is assigned to one of the players. To remove the effect of randomness of the dataset division, we run our experiments three times and only present the averages. Now, each player splits its dataset further into a training set ($80\%$) and a verification set ($20\%$). The players can run the SGD algorithm alone or together, where additional privacy mechanisms can be deployed. The accuracy of the trained model is measured via root mean square error: $RMSE=\sqrt{\frac{\sum e_{iu}^2}{|R|}}$.
    
    \vspace{-0.5cm}
    \subsection{Alone vs Together}
    \label{sec:exp_alone}
    \vspace{-0.25cm}
    
\begin{figure}[h]
        \centering
        \includegraphics[width=8cm]{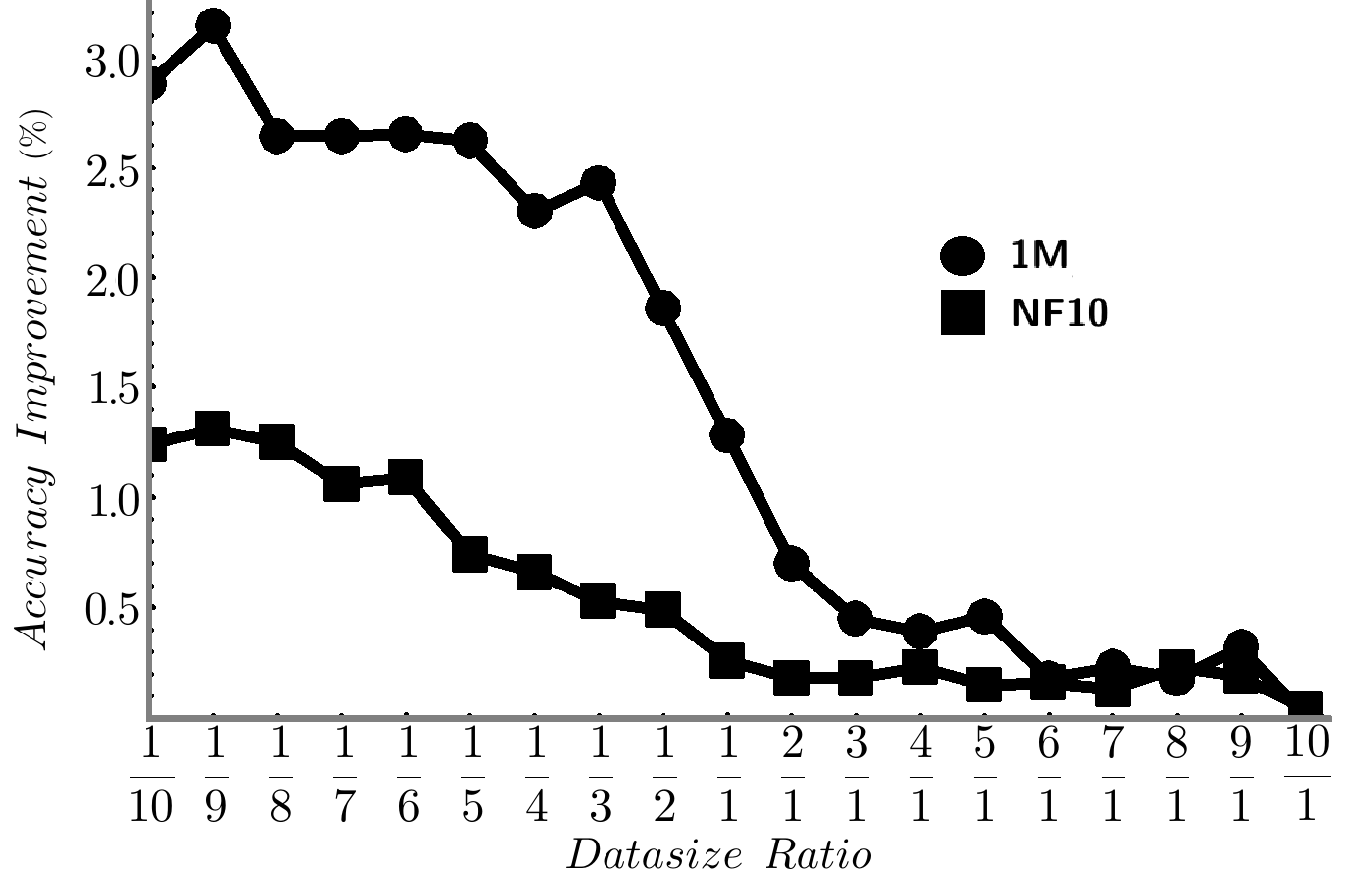}
        \caption{Accuracy improvement ($y$ axis) of training together using 1M/NF10 datasets where one player has $x$ times more (less) data than the other ($x$ axis)}
        \label{fig:size}
    \end{figure}

    First, we compare the achieved accuracy with and without the other player's data in Fig. \ref{fig:size}. The horizontal axis represents the ratio of the user-set sizes: how 1M and NF10 was split into two. More precisely, $x=\frac{\alpha}{\beta}$ represents that Player 1's dataset  is $\frac{\alpha}{\beta}$ times the size of Player 2's dataset. The vertical axis shows the normalized accuracy difference between training alone and together: $y=\frac{\theta_n-\Phi_n^M}{\theta_n}$. In other words, $y$ is the accuracy improvement via training together; $y=0$ represents the accuracy of training alone.
    
    \pagebreak
    \vspace{-0.0cm}
    \vspace{-0.25cm}
    
    It is clear that training together is superior to training alone for both datasets and  all size ratios. Fig. \ref{fig:size} also shows that the owner of the smaller dataset benefits more from collaboration; a well-expected characteristic.
    
    %From these results, we can extrapolate that a player with less data is more interested in collaboration since it's accuracy improves more. On the other hand, the player with more data can only be interested when its privacy weight is significantly lower than the accuracy weight ($C\ll B$).
    
    %In case of the 1M dataset (left), if one player's user-set is multiple times smaller than the other group's, the SGD algorithm is outperformed ($<0\%$) by the baseline prediction $\hat{r}_{ui}=0$. This indicates, if the sample size is too small, it does not represent the gradient correctly, so the algorithm does not converge to a better solution. In case of NF10 (right) this is not the case since it is a larger dataset ($\approx\times10$), i.e., SGD does improve ($>2.5\%$) the model even when the user-set is more times smaller than the other group's.
    
    \vspace{-0.5cm}
    \subsection{One Player is Privacy Concerned}
    \label{sec:exp_unc}
    \vspace{-0.25cm}
    
    Training together achieves higher accuracy than training alone. The question is, how does the situation change with a privacy mechanism in place. First we analyze the CaaS scenario introduced in Def. \ref{def:caas}. Without loss of generality, we can assume $p_2=0$. Player 1's options are either to set $p_1$ for \textit{Sup} or $\varepsilon_1$ for \textit{bDP}. 
    
    \vspace{-0.25cm}
    \begin{figure}[h]
        \centering
        \includegraphics[width=8cm]{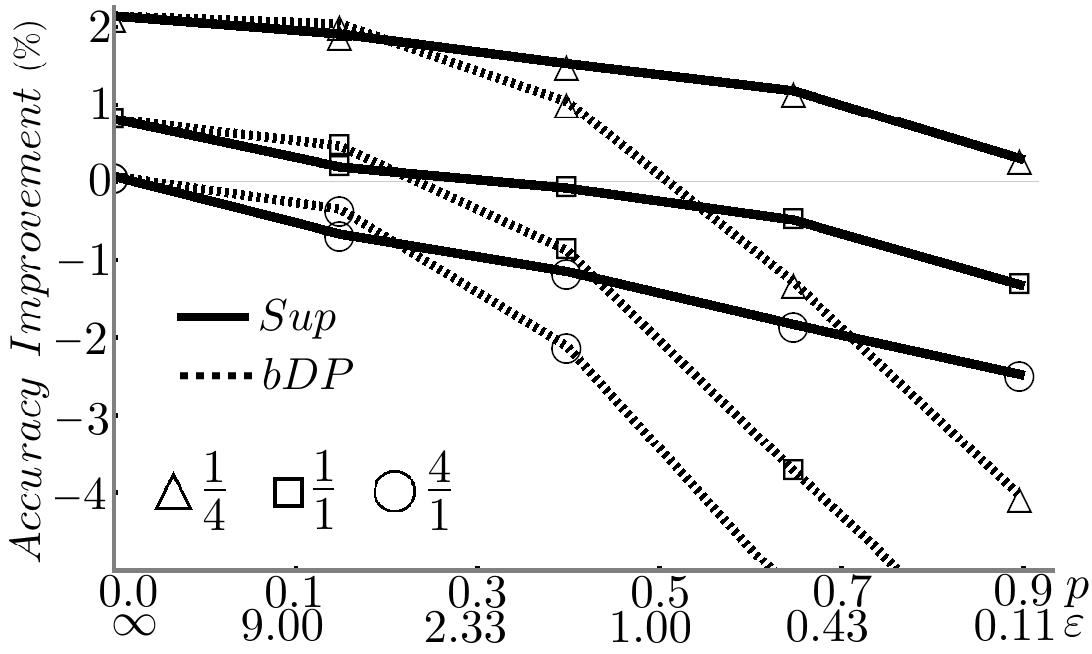}
        \caption{Accuracy improvements of training together ($y$ axis) for different privacy levels ($x$ axis) with the 1M dataset is divided such that the data size ratios are $0.25$, $1$ and $4$}
        \label{fig:ratio}
    \end{figure}
    \vspace{-0.25cm}
    
    Fig. \ref{fig:ratio} shows the tradeoff between accuracy and privacy when the 1M dataset is divided such that the data size ratio is $0.25$, $1$ and $4$ (from Player 1's perspective). The horizontal axis is the privacy parameter $p_1$ ($\varepsilon_1$) while the vertical axis shows the normalized improvement on accuracy achieved by training together (similar to Fig. \ref{fig:size}). In Fig. \ref{fig:ratio}, we can observe both higher ($y>0$) and lower ($y<0$) collaborative accuracy regions. Note, that we only show the case for 1M and select dataset size ratios as we found that using other size ratios or the NF10 dataset produce similar curves. The main observation is valid in all settings: as the dataset size ratio increases the accuracy improvement decreases.
    
    These results suggest that the realistic privacy parameters the players can apply (to obtain a more accurate model) depend on the relative size of their datasets: a player with relatively smaller dataset (e.g., triangle in Fig. \ref{fig:ratio}) can apply a stronger privacy parameter (and still obtain more accurate model) than a player with a relatively larger dataset (e.g., circle in Fig. \ref{fig:ratio}). 
    
    This finding confirms our assumption about the derivatives of $\Phi^M_n$ in Cor. 1: since the relative dataset size effects the obtained accuracy with a constant, a change in the privacy parameter effects the accuracy the same way independently of these ratios. This is indeed implies that $\partial_{p_1}\Phi_1^M=\partial_{p_2}\Phi_2^M$.
    
    \vspace{-0.5cm}
    \subsection{Both Players are Privacy Concerned}
    \vspace{-0.25cm}
    
\begin{figure}[h]
        \centering
        \includegraphics[width=8cm]{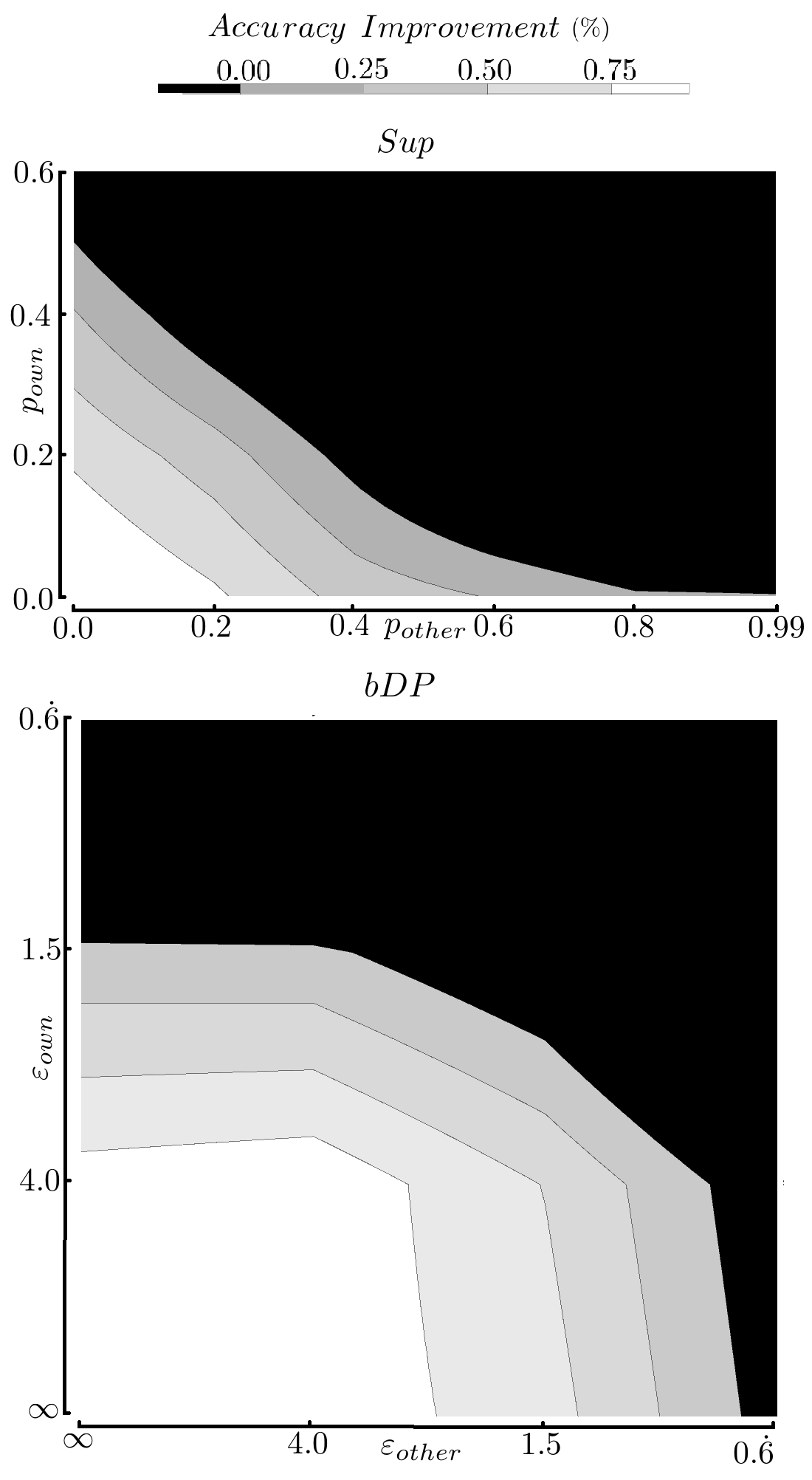}
        \caption{Accuracy improvement of collaboration (when 1M was split equally) where 0 represents the accuracy level of training alone. The applied mechanisms are \textit{Sup} and \textit{bDP}, respectively}
        \label{fig:cross}
    \end{figure}

    In this section we jointly train a model where both players employ the same mechanism $M$ with privacy level $p_1$ ($\varepsilon_1$) and $p_2$ ($\varepsilon_2$). Sec. \ref{sec:exp_alone} and \ref{sec:exp_unc} already pointed out that a player with a significantly larger dataset would not benefit much from collaboration. Consequently, for this experiment we use datasets with similar sizes. This makes our scenario symmetric, i.e., it is enough to demonstrate the accuracy change for one player. We obtained results for both 1M and NF10, but due to their similarity we only display the results for 1M in Fig. \ref{fig:cross}. We use the notation $p_{own}$ ($\varepsilon_{own}$) and $p_{other}$ ($\varepsilon_{other}$) to represent the privacy parameters from the perspective of the player under scrutiny.
    
    Fig. \ref{fig:cross} shows the normalized accuracy improvement  for different privacy parameters with privacy mechanism $M\in\{Sup,bDP\}$. It is visible that independently from $M$, accuracy is more sensitive to the player's own privacy parameter than to that of the other.
    
    These results suggest that a player's data is more useful to herself than to the other player. In other words, by degrading the quality of a given player's data (via a privacy mechanism), this player's accuracy will be affected more than the accuracy of the other player. This means, if the players would have additional incentive to undermine the other player's accuracy (e.g., competing companies), by doing so they would actually decrease their own accuracy more.
    
    \vspace{-0.25cm}
    \vspace{-0.0cm}
    
    %The more a player applies \textit{Sup}, the less its effect is on the model since Eq. (\ref{eq:upd}) is applied fewer times in the \textit{Collaborative Learning} process. Still that small effect is positive because the players are using their real ratings (i.e., without noise). As a result, the \textit{Collaborative Learning} cannot be worse than training alone for a player who hides nothing ($p_{own}=0$) even if the other player removes significant part of its dataset (e.g., $p_{other}=0.99$) since more data corresponds to better prediction (due to the definition of $\Phi$) as it can be seen on the left side of Fig. \ref{fig:cross}.
    
    %On the other hand, in case of DP the size of this effect is the same independently from the level of noise applied since Eq. (\ref{eq:upd}) is invoked the same times during Collaborative Learning. However, the more noise is applied, the less useful the effect is. Furthermore, at some point it does not even improve the accuracy of the model: the \textit{Collaborative Learning} is already less accurate for a player without noise ($\varepsilon_{own}=\infty$) than training alone if the other player apply as much as $0.6$-bDP or $2$-uDP as seen in the middle and right side of Fig. \ref{fig:cross}. Furthermore \textit{uDP} degrades the accuracy more than \textit{bDP}. This is what we expect since its privacy guarantee is strictly stronger (see Sec. \ref{sec:dp}).

	\vspace{-0.5cm}
    \section{Theory meets Practice}
    \label{sec:phi}
    \vspace{-0.25cm}
    
    Using the obtained values of $\Phi_n^M$ from Sec. \ref{sec:exp}, it is possible to interpolate the privacy-accuracy trade-off function and determine numerical NEs. In this section, we interpolate $\Phi_n^M$ for the RecSys scenario using the empirical results from Sec. \ref{sec:exp} and combine it with the results in Sec. \ref{sec:anal} to obtain exact equilibria.  
    
    \vspace{-0.5cm}
    \subsection{Interpolation via Experiments}
    \label{sec:inter}
    \vspace{-0.25cm}
    
    As an example, we set $b$ and $c$ are defined linear as in Eq. (\ref{eq:spec}). Now, we interpolate directly $b$ instead of $\Phi^M_n$, i.e., we interpolate the percentage-wise improvement difference $\frac{\theta_n-\Phi^M_n}{\theta_n}$ shown in Fig. \ref{fig:cross}. We use \textit{Mathematica}'s\footnote{\url{https://www.wolfram.com/mathematica/}} built-in $\mathsf{Interpolate}$ function with $\mathsf{InterpolationOrder\rightarrow1}$ setting in order to have a monotone approximation which is required by Def. \ref{def:b}. Via this interpolation the exact NE can be determined for the specific dataset and algorithm defined in Sec. \ref{sec:set}. In the rest of this section we calculate the precise NE when 1M dataset is split equally between the players.  
    
    \vspace{-0.5cm}
    \subsection{One Player is Privacy Concerned}
    \vspace{-0.25cm}
    
    In this CaaS scenario we assume Player $1$ is privacy unconcerned. Due to Th. \ref{th:unc}, this player's BR is $\hat{p}_1=0$. Now the utility function of Player $2$ is:
    
    \vspace{-0.25cm}
    \begin{equation}
    \label{eq:caas_ut}
    u_2(0,p_2)=B_2\cdot\overbrace{\left[\frac{\theta_2-\Phi(0,p_2)}{\theta_2}\right]^+}^{b(0,p_2)}-C_2\cdot c(p_2)
    \end{equation}
    
\begin{figure}[h]
        \centering
        \includegraphics[width=8cm]{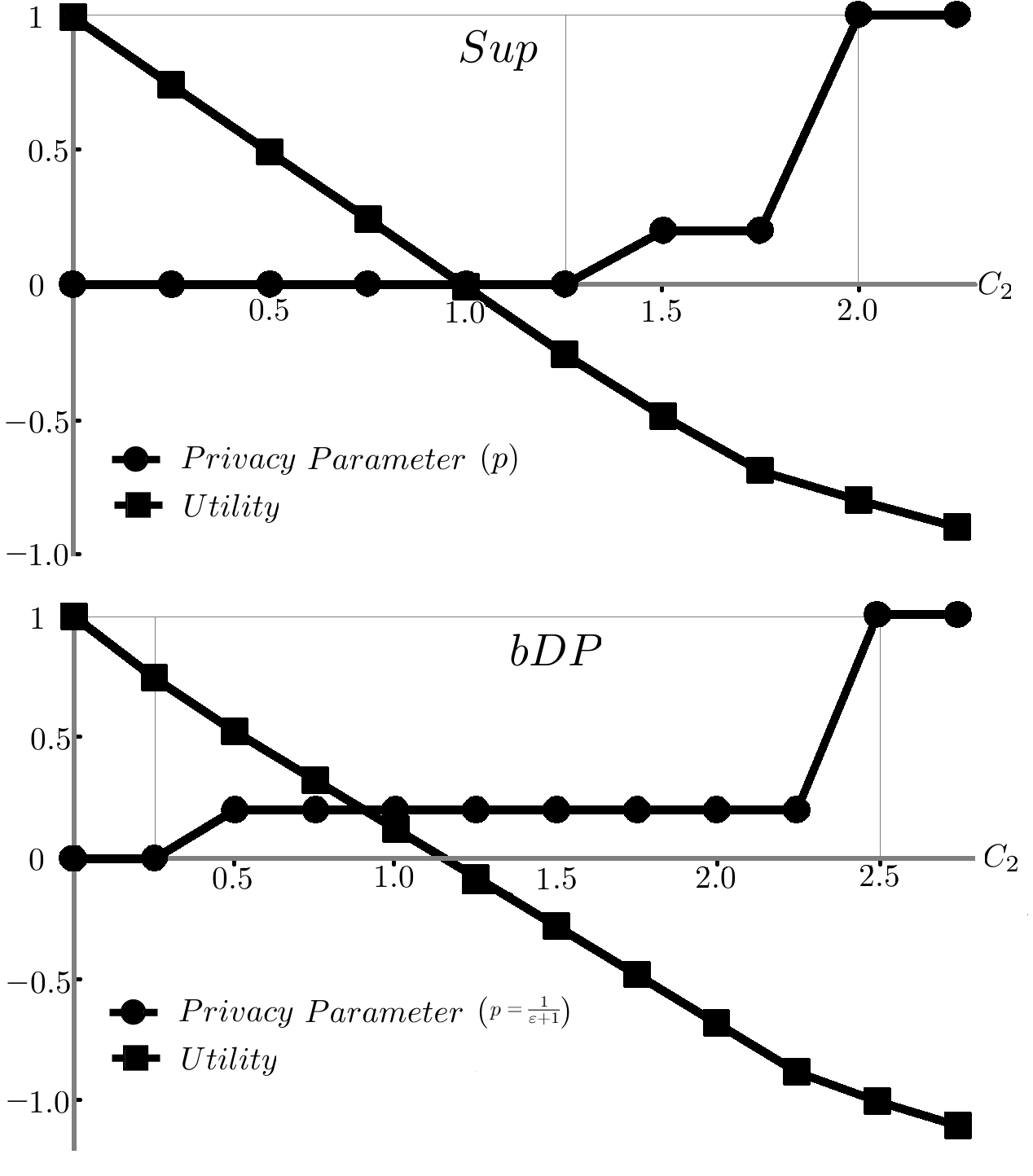}
        \caption{BR and the corresponding utility for Player $2$ when Player $1$ is privacy unconcerned}
        \label{fig:utility}
    \end{figure}

    As Lemma \ref{lemma:1} and \ref{lemma:2} states, there is a lower and upper bound on $\frac{C_2}{B_2}$ for Player $2$ which ensures that the BR $\hat{p}_2$ is either 0 or 1. We determine the exact bounds using our interpolation. Furthermore the utility of Player $2$ (Eq. (\ref{eq:caas_ut})) has to be positive as it is stated in Th. \ref{th:c_vs_uc_gen}, otherwise there is no incentive for Player $2$ to participate in the CoL process. These limits are visible Tab. \ref{tab:limit} where the right side shows the bounds corresponding to Lemma \ref{lemma:1} and \ref{lemma:2} for both privacy method while the left side corresponds to the non-negativity condition on the utility of Player $2$. These bounds on $\frac{C_2}{B_2}$ are also visible in Fig. \ref{fig:utility} where $B_2=1$.
    
    \vspace{-0.25cm}
    \begin{table}[h]
        \centering
        \begin{tabular}{|c||c||c|c|}
            \hline
            $0\leq u_2(0,\hat{p}_2)$ if & & $\hat{p}_2$ & if \\
            \hline
            \hline
            \multirow{2}{*}{$\frac{C_2}{B_2}\leq\num{0.990}$} & \multirow{2}{*}{Sup} & 0 & $\frac{C_2}{B_2}\leq\num{1.400}$ \\
            \cline{3-4}
            & & 1 & $\frac{C_2}{B_2}\geq\num{1.827}$ \\
            \hline
            \multirow{2}{*}{$\frac{C_2}{B_2}\leq\num{1.150}$} & \multirow{2}{*}{bDP} & 0 & $\frac{C_2}{B_2}\leq\num{0.349}$ \\
            \cline{3-4}
            & & 1 & $\frac{C_2}{B_2}\geq\num{2.251}$ \\
            %			\hline
            %			\multirow{2}{*}{$\frac{C_2}{B_2}\leq\num{0.990}$} & \multirow{2}{*}{uDP} & 0 & $\frac{C_2}{B_2}\leq\num{2.475}$ \\
            %			\cline{3-4}
            %			& & 1 & $\frac{C_2}{B_2}\geq\num{2.476}$ \\
            %			\hline
        \end{tabular}
        \vspace{0.1cm}
        \caption{NEs for Player $2$ when Player $1$ is privacy unconcerned}
        \label{tab:limit}
    \end{table}
    \vspace{-0.75cm}
    
    In Fig. \ref{fig:utility} we display the BR with its corresponding utility when the utility function is normalized by $B_2$ (i.e., $B_2=1$) as in the proof of Th. \ref{th:uc_vs_uc_gen}. This transformation keeps the sign, i.e., if the utility negative, the BR is not a NE, since no collaboration corresponds to higher utility. 
    
    As it is visible, in case of \textit{Sup} the interval defined by the two lemmas (represented by the two vertical thin gray line) are corresponding to negative utility, so for this privacy mechanisms the NE is either collaboration without privacy protection or no collaboration, depending on the weight ratio. More precisely, according to Th. \ref{th:c_vs_uc_gen} the NE is $(p_1^*,p_2^*)=(0,0)$ if $\frac{C_2}{B_2}\leq b(\theta_2,\Phi_2(0,0))=\num{0.990}$ and $(p_1^*,p_2^*)=(1,1)$ otherwise. %Note, that the highest benefit possible which is the percentage-wise improvement on the accuracy is almost $1\%$.
    
    In case of \textit{bDP}, some part of the interval created by the lemmas corresponds to positive utility, i.e., there exists a non-trivial NE. More precisely, if $\num{0.349}\leq\frac{C_2}{B_2}\leq\num{1.150}$ then $p_2^*$ is neither $0$ nor $1$. Note, that the BR function is step-like because of the piecewise linear interpolation. As such, within this interval the NE is $p_2^*=0.2\Leftrightarrow\varepsilon_2^*=4$. The \textit{Prive of Privacy} of this NE is $PoP(0,0.2)=\num{0.066}$, so due to privacy concerns less than $7\%$ of the overall achievable accuracy is lost. 
    
    \vspace{-0.25cm}
    \vspace{-1cm}
    \subsection{Both Player are Privacy Concerned}
    \vspace{-0.25cm}
    
    In this section we will focus on \textit{bDP} privacy mechanism. Th. \ref{th:uc_vs_uc_gen} states that when both player is privacy concerned a non-trivial NE exists. To find it, we will use BR dynamics \cite{harsanyi1988general}. This will eventually converge to a NE due to the following theorem:
    
    \vspace{-0.25cm}
    \begin{theorem*}[Monderer \& Shapley \cite{monderer1996potential}]
        In a finite\footnote{As the CoL game is not finite, we discretized the actions spaces of the players with floating point numbers.} potential game, from an arbitrary initial outcome, the BR dynamics converges to a pure strategy NE.
    \end{theorem*}
    \vspace{-0.25cm}
    
    In BR dynamics the players update their strategies in the next round based on the their BRs to the strategy what the other player played last round. We start the iteration from $(p_1,p_2)=(0,0)$\footnote{The higher the privacy level from where the BR dynamics starts, the bigger the interval of weight ratios in which case it converges to 1 (i.e., no collaboration).} and update the players's strategies alternately starting with Player $1$. The NEs where the process converged are visible in Tab. \ref{tab:exact} with the corresponding \textit{Price of Privacy} values for discrete weight ratios $\{0,0.1,\dots\}$.
    
    \vspace{-0.25cm}
    \begin{table}[h]
        \centering
        \begin{tabular}{|c||c|c|c|}
            \hline
            $\frac{C_2}{B_2}\in$ $\rightarrow$ & $[0,0.3]$ & $[0.4,0.9]$ & $[1,\infty]$ \\
            \hline
            \hline
            \multirow{2}{*}{$\frac{C_1}{B_1}\in[0,0.3]$} & $(0,0)$ & $(0,0.2)$ & $(1,1)$ \\
            & $\num{0.000}$ & $\num{0.066}$ & $\num{1.000}$ \\
            \hline
            \multirow{2}{*}{$\frac{C_1}{B_1}\in[0.4,0.9]$} & $(0.2,0)$ & $(0.2,0.2)$ & $(1,1)$ \\
            & $\num{0.066}$ & $\num{0.131}$ & $\num{1.000}$ \\
            \hline
            \multirow{2}{*}{$\frac{C_1}{B_1}\in[1,\infty]$} & $(1,1)$ & $(1,1)$ & $(1,1)$ \\
            & $\num{1.000}$ & $\num{1.000}$ & $\num{1.000}$ \\
            \hline
        \end{tabular}
        \vspace{0.1cm}
        \caption{NEs for different weight ratios}
        \label{tab:exact}
    \end{table}
    \vspace{-1cm}
    
    These approximated result suggests that players with low privacy weight prefers to train together without any protection while high privacy weight ensures no collaboration. Furthermore, the narrow interval in-between corresponds to collaboration with very limited privacy protection (e.g., $\varepsilon=4$) or no collaboration at all. 

    \vspace{-0.5cm}
    \section{Approximating $\Phi$ in Practice}
    \label{sec:approx}
    \vspace{-0.25cm}
    
    Direct interpolation is only possible when both datasets are fully available. In a real-world scenario $\Phi^M_n$ must be approximated by other means. In this section, besides a short overview of the whole process before collaboration, we demonstrate a simple approach to fill the gap in the CoL game caused by the obscurity of $\Phi^M_n$. 
    
    Note, that our intention is not to provide a sound method to approximate the effect of privacy mechanisms on the accuracy of complex training algorithms, but rather to show a direction how it could be done. More research is required in this direction, and we consider this to be an interesting future work.
    
    \vspace{-0.5cm}
    \subsection{Heuristic Parameters}
    \vspace{-0.25cm}
    
    As we argued in Sec. \ref{sec:exp}, we can assume that the player's datasets are from similar distributions, i.e., the players can imitate CoL by mimicking the player's datasets via splitting their own datasets into two and approximate $\Phi_n^M$ locally. 
    
    Based on empirical results, we establish a heuristic formula which minimizes the error of this local approximation based on the size and density of the players' datasets. The formula can be seen in Eq. (\ref{eq:heuristic}) where $d$ is the density of the datasets and $D_n$ is player $n$'s dataset. The details of the related experiment can be found in App. \ref{app:self}. We refer to the true privacy-accuracy trade-off function as $\Phi_n^M$ and our approximation via self-division as $\widetilde{\Phi_n^M}$.
    
    \vspace{-0.25cm}
    \begin{equation}
    \label{eq:heuristic}
    \num{100000}\approx d\cdot|D_n|=\frac{|D_n|^2}{|U_n|\cdot|I|}
    \end{equation}
    
    %In more details, in Appendix \ref{app:self} we established heuristic connection between the dataset size and the difference between the true and the approximated accuracy improvement if the density is fixed. As such, self division could be used with other density/dataset sizes also with some adjustment/scaling. As we do not explored this direction, for the sake of this example we set the parameters of the datasets to fulfill Eq. (\ref{eq:heuristic}).
    
    \vspace{-0.5cm}
    \subsection{The Whole Process Before Collaboration}
    \vspace{-0.25cm}
    
\begin{figure}[h!]
        \centering
        \includegraphics[width=7cm]{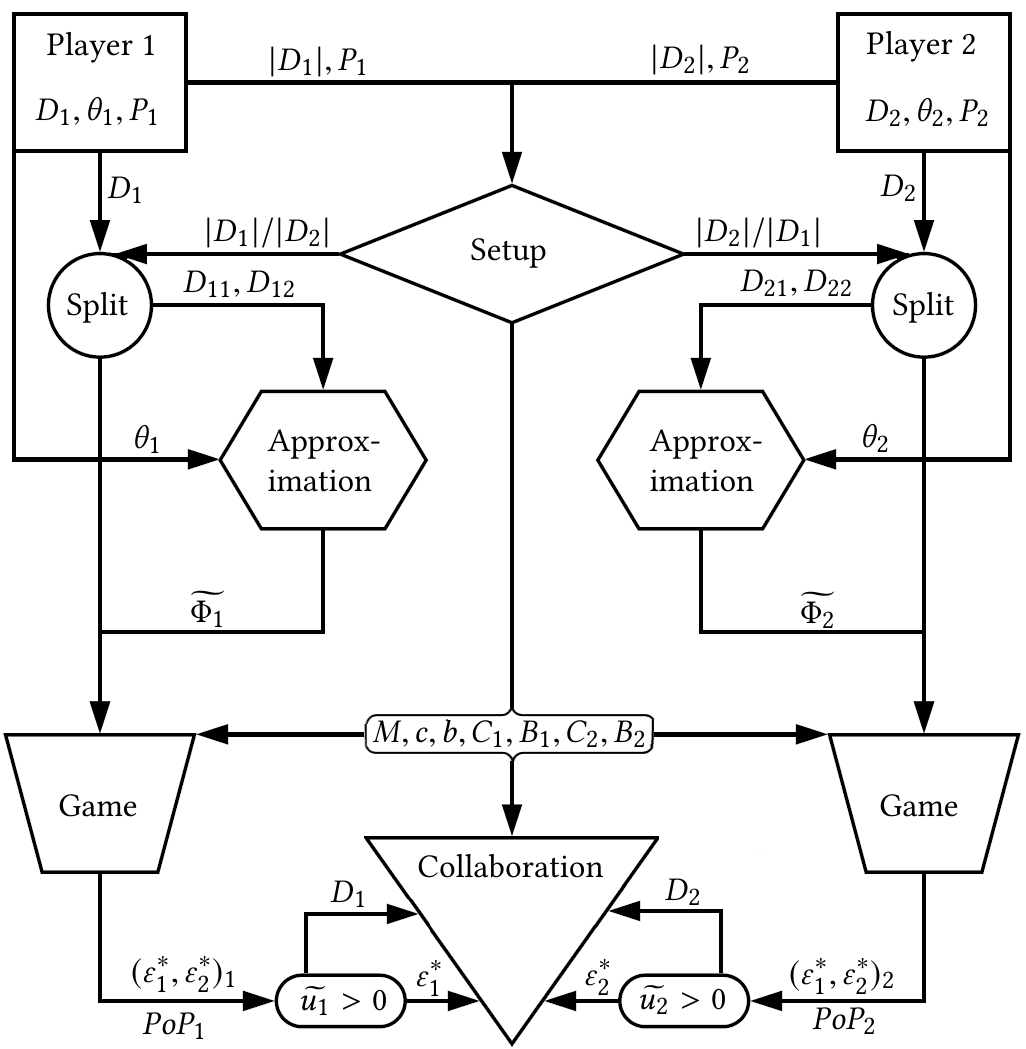}
        \caption{The process diagram which describes what are the steps of the players before collaboration}
        \label{fig:summ}
    \end{figure}

    In this section we show how our game theoretic model can be used from scratch. That is, the players have two datasets and they would like to know whether to train together and with what privacy parameter. These questions can be answered with the help of the CoL game, but first its parameters must be established. A process diagram describing the whole process is presented in Fig. \ref{fig:summ}.
    
    \vspace{-0.25cm}
    \vspace{-0.25cm}
    
    \begin{itemize}
        \item \textbf{Initialization}: The players have their datasets $D_n$ with the corresponding privacy policies $P_n$ and the accuracy they achieve by training alone $\theta_n$.
        \item \textbf{Setup}: Based on the size of the datasets $|D_n|$ and the privacy policies $P_n$, the players determine which privacy preserving method $M$ to use, what benefit $b$ and cost $c$ function to apply, and what should be the corresponding weight parameters $C_1,B_1,C_2,B_2$.
        \item \textbf{Split}: Based on the dataset size ratio $|D_n|/|D_m|$, the players split their own datasets into two $D_{n1},D_{n2}$ to mimic the original datasets. 
        \item \textbf{Approximation}: Based on the newly created datasets $D_{n1},D_{n2}$ and the accuracy obtained by training alone $\theta_n$, the players approximate the accuracy improvement of training together $\widetilde{\Phi_n}$.
        \item \textbf{Game}: Using the approximated privacy-accuracy trade-off function $\widetilde{\Phi_n}$ and $M,b,c,C_1,B_1,C_2,B_2$ determined by the setup phase, the players determine the NE $(\varepsilon_1^*,\varepsilon_2^*)_n$ and its corresponding \textit{Price of Privacy} value via the CoL game.
        \item \textbf{Collaboration}: If the approximation suggest that training together is beneficial for both participant (i.e., $\forall n\in\{1,2\}:\widetilde{u_n}(\varepsilon_1^*,\varepsilon_2^*)>0$), than they collaborate using their dataset $D_n$ with the approximated optimal privacy parameter $\varepsilon_n^*$.
    \end{itemize}
    \vspace{-0.25cm}
    
    As we focused mainly on the ``Game'' step and partially on the ``Approximation'' step, we manually chose the parameters determined by the ``Setup'' phase. In this example we choose \textit{bDP} as privacy-preserving mechanism $M$. We assume that players have a chunk of the preprocessed NF rating dataset, which contains only movie ratings. As such, it is expected that the players value privacy less than accuracy: for the sake of this example, we set $B_1=B_2=1$ and $C_1=C_2=0.1$. We use the benefit and privacy loss functions defined in Eq. (\ref{eq:spec}).
    
    As we argued previously, self-division is the most punctual when Eq. (\ref{eq:heuristic}) holds. Since the density of the original NF dataset is $d\approx0.01$, we assign both players datasets with 10 million ratings: we randomly choose $20\%$ of the users from NF10 and assign them to either one of the players.
    
    The players separately approximate $\Phi_n$ by self-division, therefore, $\widetilde{\Phi_1}$ and $\widetilde{\Phi_2}$ are not necessarily the same. The exact values of these approximations can be seen in App. \ref{app:excel} together with the true value of $\Phi_n$.\footnote{Note, that $\Phi_n$ itself is interpolated from its actual value at measured points.} We found that the RMSE of $\widetilde{\Phi_n}$ is around $0.001$ for both players.%, thus Eq. (\ref{eq:heuristic}) seems to hold for datasets with different densities/sizes as well. 
    
    Using $\widetilde{\Phi_1}$ Player 1 approximates the NE as $(\widetilde{p_1^*},\widetilde{p_2^*})=(0,0)$ while Player 2 reaches the same conclusion via $\widetilde{\Phi_2}$. This means \textit{Price of Privacy} is zero. The approximated utilities are $\widetilde{u_1}=\num{0.18}$ and $\widetilde{u_2}=\num{0.07}$ respectively. The actual utilities in case of $(\widetilde{p_1^*},\widetilde{p_2^*})$ are $(0.21,0.07)$, which are very close to the approximated values. While utility approximation is fairly accurate, $\Phi_n$ actually corresponds to a slightly different NE: $(p_1^*,p_2^*)=(0.2,0.2)$ with utility $(u_1,u_2)=(0.14,0.06)$ and $PoP=0.25$. Note, that while both players obtain a higher payoff via $\widetilde{p_n^*}$ that is not an actual NE.

    \vspace{-0.5cm}
    \section{Related Work}
    \label{sec:related}
    \vspace{-0.25cm}
    
    We divide related literature into two groups based on the two main topic of this paper: distributed ML and GT.
    
    \vspace{-0.5cm}
    \subsection{Distributed Machine Learning}
    \vspace{-0.25cm}
    
    ML is frequently implemented in a distributed fashion for efficiency reasons. To tackle its emerging privacy aspect, Privacy Preserving Distributed ML was introduced, where the locally trained models are safely aggregated.
    
    Distributed training scenarios unanimously assume a large number of participants and the involvement of a third party such as in \cite{pathak2010multiparty,rajkumar2012differentially,hamm2016learning,mcmahan2016communication,pawlick2016stackelberg}. In more details, in \cite{pathak2010multiparty} mutually untrusted parties train classifiers locally and aggregate them with the help of an untrusted curator. In the introduced $\varepsilon$-DP protocol, achieved accuracy depends on the number of parties and the relative fractions of data owned by the different parties. In \cite{rajkumar2012differentially} these dependencies were eliminated for a SGD training algorithms. On the other hand, authors used $(\varepsilon,\delta)$-DP, a weaker form of DP.
    
    More recently in \cite{hamm2016learning} an $\varepsilon$-DP classifier was introduced with error bound $O((\varepsilon N)^{-2})$ compared to the result of the non-private training where $N$ is the number of participants. This approach results in strong privacy guarantees without performance loss for large $N$. Federated Learning introduced in \cite{mcmahan2016communication} follows another approach, where the users generate pairwise noise to mask their data from the aggregator. The bottleneck of this approach is the communication constraints. Furthermore, the solution is not applicable to two participants.
    
    All these works assumed the existence of a third-party aggregator; however, in our work the data holders themselves train a model together to achieve higher accuracy than what they would have obtained if training in isolation. Furthermore, all of these works are neither suitable nor efficient for two participants.
    
    \vspace{-0.5cm}
    \subsection{Game Theory}
    \vspace{-0.25cm}
    
    In \cite{pawlick2016stackelberg} the learning process was modeled as a Stackelberg game amongst $N+1$ players where a learner declares a privacy level and then the other $N$ data holders respond by perturbing their data as they desire. The authors concluded that in equilibrium each data holder perturbs its data independently of the others, which leads to high accuracy loss. 
    
    The closest to our work are \cite{ioannidis2013linear,chessa2015game,wu2017game}. In \cite{ioannidis2013linear} a linear regression scenario was studied where the features were public but the data were private. With these settings, the authors proved the existence of a unique non-trivial NE, and determined its efficiency via the Price of Stability.
    
    A simpler problem was modeled in \cite{chessa2015game}: estimating a population's average of a single scalar quantity. The authors studied the interaction between agents and an analyst, where the agents can either deny access to their private data or decide the level of precision at which the analyst gets access. Findings include that it is always better to let new agents enter the game as it results in more accurate estimation, and the accuracy can further be improved if the analyst sets a minimum precision level.
    
    In both previous scenarios, players would like to learn a model which represents the whole population. The accuracy of the estimate is a public good (i.e., non-exclusive and non-rival \cite{harsanyi1988general}). On the contrary, in CoL the players seek to selfishly improve their own accuracy as that is in their own self-interest. As such, they measure the accuracy of the trained model by how well it fits to their own datasets, which can result in different accuracy levels. Furthermore, these works focused on particular tasks (linear regression and scalar averaging) while our model is applicable for any training mechanism.
    
    \cite{wu2017game} studied the problem of private information leakage in a data publishing scenario where datasets are correlated. As such, the utility function for an agent consists of the benefit of publishing its own sanitized dataset and the privacy leakage which depends on the privacy parameters of all involved agents. Opposed to this, in our model the datasets are independent while the benefit is affected by all the players' actions. Thus, the accuracy of the training depends on the privacy parameters of both agents, while the privacy loss depends only on the privacy parameter of a single agent.

    \vspace{-0.5cm}
    \section{Conclusion}
    \label{sec:conc}
    \vspace{-0.25cm}
    
    In this paper, we designed a Collaborative Learning process among two players. We defined two player types (privacy concerned and unconcerned) and modeled the training process as a two-player game. We proved the existence of a Nash Equilibrium with a natural assumption about the privacy-accuracy trade-off function ($\Phi$) in the general case, while provided the exact formula when one player is privacy unconcerned. We also defined \textit{Price of Privacy} to measure the overall degradation of accuracy due to the player's privacy protection. 
    
    On the practical side, we studied a Recommendation System use case: we applied two different privacy-preserving mechanisms (suppression and bounded differential privacy) on two real-world datasets (MovieLens, Netflix). We confirmed via experiments that the assumption which ensures the existence of a Nash Equilibrium holds. Moreover, as a complementary work besides the designed game, we interpolated $\Phi$ for this use case, and devised a possible way to approximate it in real-world scenarios. Our main findings are:
    
    \begin{itemize}
        \item Privacy protection degrades the accuracy heavily for its user.
        \item Collaborative Learning is practical when either one player is privacy unconcerned or the players have similar dataset sizes and both players' privacy concerns (weights) are relatively low.
    \end{itemize}
    \vspace{-0.25cm}
    
    %In relation with the research questions in Sec. \ref{sec:int}, we explicitly answered $Q_1$ via Fig. \ref{fig:cross} for a specific learning algorithm, datasets and privacy preserving method. We answered $Q_2$ via Th. \ref{th:unc}, \ref{th:c_vs_uc_gen} and \ref{th:uc_vs_uc_gen}. Furthermore, by the measurement defined in Def. \ref{def:pop} we provided a way to directly answer $Q_3$.
    
    %\subsection{Future Works}
    
    \textbf{Future work. }There are multiple opportunities to improve this line of work such as upgrading the CoL process by controlling the other party's updates. Another possibility is to design a repetitive game where each player faces a decision after each iteration or make the game asymmetric by defining the weights $B$ and $C$ in private. %Introducing more parameters such as time or electricity consumption or 
    Incorporating the impact of the potential adversarial aspect for competing companies, and thus investigating a more elaborate utility function is another intriguing possibility. Finally, as complementary work, how to determine the weight parameters for specific scenarios and how to approximate $\Phi$ is crucial for the usability of the model in the real world.
    
    %\section{Acknowledgment}
    
    %	We would like to thank the anonymous reviewers for their deep and meaningful comments. 

    \vspace{-0.5cm}
	\bibliography{REF}{}
	\bibliographystyle{alpha}
	
    \newpage
    
	\begin{appendices}
        \section{List of Abbreviations}
        \label{app:abr}
        
        \begin{table}[h]
            \centering
            \begin{tabular}{|c|l|}
                \hline
                \bf{Abr.} & \bf{Meaning}\\
                \hline
                \hline
                $ML$ & Machine Learning\\
                \hline
                $MF$ & Matrix Factorization\\
                \hline
                $SGD$ & Stochastic Gradient Descent\\
                \hline
                $DP$ & Differential Privacy\\
                \hline
                $Sup$ & Suppression\\
                \hline
                $RecSys$ & Recommender System\\
                \hline
                $CoL$ & Collaborative Learning\\
                \hline
                $PoP$ & Price of Privacy\\
                \hline
                $GT$ & Game Theory\\
                \hline
                $NE$ & Nash Equilibrium\\
                \hline
                $BR$ & Best Response\\
                \hline
            \end{tabular}
            \vspace{0.1cm}
            \caption{Frequently used abbreviations}
        \end{table}
        
        \section{Proofs for Sec. \ref{sec:anal}}
        \label{app:proofs}
        
        \begin{proof}[Proof Th. \ref{th:tri}]
            Without loss of generality, assume Player $2$ sets $p_2=1$. The highest accuracy Player $1$ can achieve corresponds to $p_1=0$ due to the Def. \ref{def:b} and \ref{def:phi}. From Def. \ref{def:phi} we can also deduce that if one player sets its privacy parameter to maximal 1 then neither of the players can obtain higher accuracy by training together than training alone. As such, the highest accuracy what Player 1 can reach by training together when Player 2 sets its privacy parameter to maximum is less than what it would achieve by training alone: $\Phi_1(0,1)\geq\theta_1$. Note that $\Phi$ and $\theta$ measures the error, i.e., the higher these values are, the less accurate the corresponding model is.
            
            As such, $p_1=0$ does not correspond to positive benefit but only results in privacy loss. Hence, the highest payoff Player $1$ can reach is $0$ corresponding to maximal privacy protection $p_1=1$. In other words, if Player 2 sets $p_2=1$ the BR of Player 1 is also to set $p_1=1$. Since this is also true on the other way around, $(p_1^*,p_2^*)=(1,1)$ is indeed a NE which is equivalent to the case of training alone.
        \end{proof}
        
        \begin{proof}[Proof L. \ref{lemma:1}]
            If $\alpha_n=0$, the utility function in Eq. (\ref{eq:ut}) is reduced to $u_n=B_n\cdot b(\theta_n,\Phi_n)$ since $C_n=0$. This is strictly positive by definition. Also by definition $b$ is monotone decreasing in $p_n$. As a result, the utility is highest when no privacy protection is in place. As such, indeed exists $\alpha_n$ such that $\hat{p}_n=0$ is the BR for player $n$.
        \end{proof}

        \begin{proof}[Proof L. \ref{lemma:2}]
            Without loss of generality we assume $n=1$. We show that $\max_{p_1}u_1(p_1,p_2)=u_1(1,p_2)=0$ if $C_1\rightarrow\infty$ which is equivalent with the statement in Lemma \ref{lemma:2}:
            \begin{equation}
            \begin{gathered}
            \lim_{C_1\rightarrow \infty}u_1(p_1,p_2)=\\
            \lim_{C_1\rightarrow \infty}B_1\cdot b(\theta_1,\Phi_1(p_1,p_2))-c(p_1)\cdot C_1\leq\\
            \lim_{C_1\rightarrow \infty}B_1\cdot b(\theta_1,\Phi_1(0,0))-c(p_1)\cdot C_1=\\
            \lim_{C_1\rightarrow \infty}\beta_0-c(p_1)\cdot C_1=\left\{
            \begin{tabular}{ccc}
            $\beta_0$ & if & $c(p_1)=0$\\
            $-\infty$ & if & $c(p_1)>0$\\
            \end{tabular}\right.
            \end{gathered}
            \end{equation}
            As a result, $u_1(p_1,p_2)\leq\beta_0$ for some $\beta_0\geq0$ and it can only be non-negative if $c(p_1)=0$ which corresponds to $p_1=1$. The utility is maximal in this case, thus, $\max_{p_1}u(p_1,p_2)=u(1,p_2)$ which is indeed 0.
        \end{proof}
        
        \begin{proof}[Proof Th. \ref{th:unc}]
            In the proof of Lemma \ref{lemma:1} we set $C_n=0$ in which case player $n$'s BR was indeed $\hat{p}_n=0$. For more details read the proof of Lemma \ref{lemma:1}.
        \end{proof}
        
        \begin{proof}[Proof Th. \ref{th:c_vs_uc_gen}]
            The utility function $u(p_1)$ is maximal in the interval $[0,1]$ either on the border or at a point where its derivative is zero. The derivative of Eq. (\ref{eq:ut}) is
            \begin{equation}
            \label{eq:decD_gen}
            u'(p_1)=B_1b'(p_1)\Phi'(p_1)-C_1c'(p_1)
            \end{equation}
            which is zero at $\tilde{p}_1$ if 
            
            \begin{equation}
            u'(\tilde{p}_1)=0\Rightarrow\frac{b'(\tilde{p}_1)\Phi'(\tilde{p}_1)}{c'(\tilde{p}_1)}=\frac{C_1}{B_1}
            \end{equation}
            Of course the extreme point $\tilde{p}_1$ must be in $[0,1]$ and $u(\tilde{p}_1)>0$. Furthermore, this extreme point is only maximum when the second derivative is negative.
            
            On the other hand, if Eq. (\ref{eq:decD_gen}) is never zero on $[0,1]$ or the second derivative is positive at that point, the maximum of $u(p_1)$ is on the edge of the interval $[0,1]$. $u(1)=0$ since both the benefit and the privacy loss function is zero at $p_1=1$. As a result, $p_1=0$ is the maximum point if $u(0)>0$. This is indeed the case when the maximal benefit $b(0)$ is higher than the ratio of the privacy and accuracy weight $\frac{C_1}{B_1}$ as it is shown below:
            
            \begin{equation*}
            0<u(0)=B_1b(0)-C_1\Rightarrow b(0)>\frac{C_1}{B_1}
            \end{equation*}
        \end{proof}
        
        \begin{proof}[Proof Th. \ref{th:uc_vs_uc_gen}]
            We divide $u_n$ by $B_n$: $\tilde{u}_n=\frac{u_n}{B_n}$. This new function inherits the properties of $u_n$ (such as the sign, monotonicity, maximum/minimum points, etc.). As a result, a similar game with utility function $\tilde{u}_n$ has the same equilibria. Furthermore, this similar game is a potential game if the mixed second order partial derivative of the utility functions are equal. Due to the constitution of $\tilde{u}_n$, this condition is equivalent to
            \begin{equation*}
            \partial_{p_1}\partial_{p_2}b(\theta_1,\Phi_1(p_1,p_2))=\partial_{p_1}\partial_{p_2}b(\theta_2,\Phi_2(p_1,p_2))
            \end{equation*}
            This formula can be transformed into the one in the theorem by applying the chain rule for higher dimensions.
        \end{proof}
        \begin{proof}[Proof Col. 1]
            The left side of the equation in Th. \ref{th:uc_vs_uc_gen} is zero since we assumed $\partial_{p_1}\Phi_1=\partial_{p_2}\Phi_2$. On the right side $\partial_{p_1}\partial_{p_2}\Phi_2=\partial^2_{p_1}\Phi_1$ and $\partial_{p_2}\partial_{p_1}\Phi_1=\partial^2_{p_2}\Phi_2$ for the same reason. This means Th. \ref{th:uc_vs_uc_gen} holds since both sides of the equation are 0.
        \end{proof}
        
        \section{Proof of Theorem \ref{th:sens}}
        \label{app:proof2}
        
        \begin{proof}[Proof Th. \ref{th:sens}]
            Since the user set of the players are disjoint while the item set are shared, the only thing the players need to share is the item feature matrix $Q$. The effect of a single update is shown in Eq. (\ref{eq:upd}). We assume that the data points are independent, hence, the sensitivity $\tilde{S}$ of one update is
            \begin{equation*}
            \begin{gathered}
            \tilde{S}=\max_{r_{ui}}|q'_{ki}-q_{ki}|=\max_{r_{ui}}[\gamma(e_{ui}p_{uk}-\lambda q_{ki})] \\
            \leq\gamma(\Delta rp_{max}+\lambda q_{max})\\
            \end{gathered}
            \end{equation*}
            where
            \begin{itemize}
                \item $k\in[1,\kappa]$
                \item $\Delta r$ is the maximal distance of two ratings: $\Delta r=\max{r_{ui}}-\min{r_{ui}}$
                \item $p_{max}$ and $q_{max}$ are the maximal absolute value of the user and item features respectively.
            \end{itemize}
            
            $\tilde{S}$ is the sensitivity of updating a single feature, thus, to capture the full effect of the update on the vector $q_i$, we need to multiply $\tilde{S}$ with the $q_i$'s dimension $\kappa$. Moreover, we have only considered the effect of a rating on $Q$ within one iteration. However, this occurs $\iota$ times. Thus, to achieve $\varepsilon$-DP, we need to apply $\frac{\kappa\cdot\iota\cdot\tilde{S}}{\varepsilon}$ level of Laplacian noise on the ratings before the training due to the Composition Theorem. Therefore, the overall sensitivity is indeed bounded by the formula in the theorem.
        \end{proof}
        
        \section{Preprocessing}
        \label{app:pre}
        
        \begin{itemize}
            \item[1] Remove items/users with less than $10$ ratings.%\footnote{In \cite{friedman2016differential}, the authors create `fake' ratings to stabilize the item/user averages. Instead of this inflation, we chose to remove them.}.
            \item[2] For each remaining items, calculate the average rating and discount it from the corresponding $r_{ui}$'s:
            \begin{equation*}
            r'_{ui}:=r_{ui}-I_{Avg}(i)
            \end{equation*}
            \item[3] For each remaining user, calculate the average rating and discount it from the corresponding $r'_{ui}$'s:
            \begin{equation*}
            r''_{ui}:=r'_{ui}-U_{Avg}(u)=r_{ui}-I_{Avg}(i)-U_{Avg}(u)
            \end{equation*}
            \item[4] The discounted ratings as well as the averages are clamped:
            \begin{itemize}
                \item $I_{Avg}(i)\in[\min(r_{ui}), \max(r_{ui})]=[1,5]$
                \item $U_{Avg}(u)\in[-2, 2]$
                \item $r''_{ui}\in[-2, 2]$
            \end{itemize}
        \end{itemize}
        
        %\section{Unbounded Differential Privacy}
        %\label{app:uDP}
        
        %	To achieve $\varepsilon$-DP, first the ratings need to be preprocessed as described in Appendix \ref{app:pre}. Then the following process needs to be applied:
        
        %	\begin{itemize}
        %		\item[1] Set all missing ratings to 0.
        %		\item[2] Add noise to all ratings (fake and original) as it is shown in Eq. (\ref{eq:bdp}).
        %		\item[3] Choose $\delta\in\mathbb{R}$ and delete ratings in the range $[-\delta,\delta]$ as it is shown in Eq. (\ref{eq:udp}) where $\bot$ represents the removed rating.
        %		\item[4] Choose $\mu\in[0,1]$ and delete $\mu$ percentage of ratings randomly where $r_{ui}=\pm r_{max}$ as it is shown in Eq. (\ref{eq:udp}) where $\upsilon_{ui}\sim\mathbb{U}(0,1)$ is a uniform random variable.
        %	\end{itemize}
        
        %	\begin{equation}
        %		\label{eq:udp}
        %		r''_{ui}:=\left\{
        %		\begin{tabular}{lll}
        %			$\bot$ & if & $|r'_{ui}|\leq\delta$\\
        %			$\bot$ & if & $r'_{ui}=\pm r_{max}$\text{ and }$\upsilon_{ui}\geq\mu$\\
        %			$r'_{ui}$ & & otherwise\\
        %		\end{tabular}\right.
        %	\end{equation}
        
        %	Step 3 and 4 are necessary. Otherwise, the learning would be infeasible due to the rating matrix inflation at Step 1: in case of high $\varepsilon$ (low noise level), step 3 ensures that the majority of the ratings are removed. On the other hand, in case of low $\varepsilon$ (high noise level), a significant portion of the ratings are removed due to Step 4.
        
        \section{Self-Division: Experiment}
        \label{app:self}
        
        \begin{itemize}
            \item[1] We create datasets with approximately the same density but with different size:
            
            \begin{itemize}
                \item 1M: We modify the size of the dataset while keeping its density: we randomly removing users such that the remaining dataset has 1000k/800k/600k ratings (i.e., the players have 500k-500k/400k-400k and 300k-300k ratings).
                \item NF10D: We create a new dataset originated from NF10 by increasing its density to the level of 1M via filtering out the less rated items\footnote{We remove the items which have less than 250 ratings}. Afterwards we modify the size of this dataset while keeping its newly acquired density: we randomly removing users such that the remaining dataset has 8m/6m/4m/2m ratings.
            \end{itemize}
            
            \item[2] We execute CoL with $p_i\in\{0,0.2,0.4,0.6\}$ for $i=\{1,2\}$ and for $M\in\{Sup,bDP\}$ using the newly created datasets (e.g., the players have 300k/400k/$\dots$/3m/4m ratings) and we obtain the normalized accuracy improvement for both player: $\Phi'_i=\frac{\theta_i-\Phi_i^M}{\theta_i}$. 
            
            \item[3] We execute CoL with the same privacy parameters and methods using only one player's data: the players imitate CoL by halving their own datasets (e.g., the datasets sizes are 150k-150k/$\dots$/2m-2m) and we obtain $\widetilde{\Phi'}=\frac{\widetilde{\theta_n}-\widetilde{\Phi_n^M}}{\widetilde{\theta_n}}$ where $\widetilde{\theta_n}$ and $\widetilde{\Phi_n}$ corresponds to the average of the accuracies using the two half of player $n$'s data.
            
            \item[4] We calculate the RMSE between the original normalized accuracy $\Phi'$ and the approximated normalized accuracy via self-division $\widetilde{\Phi'}$ for both players and privacy methods. 
        \end{itemize}
        
        \nopagebreak
        
        \begin{figure}[h]
            \centering
            \includegraphics[width=8cm]{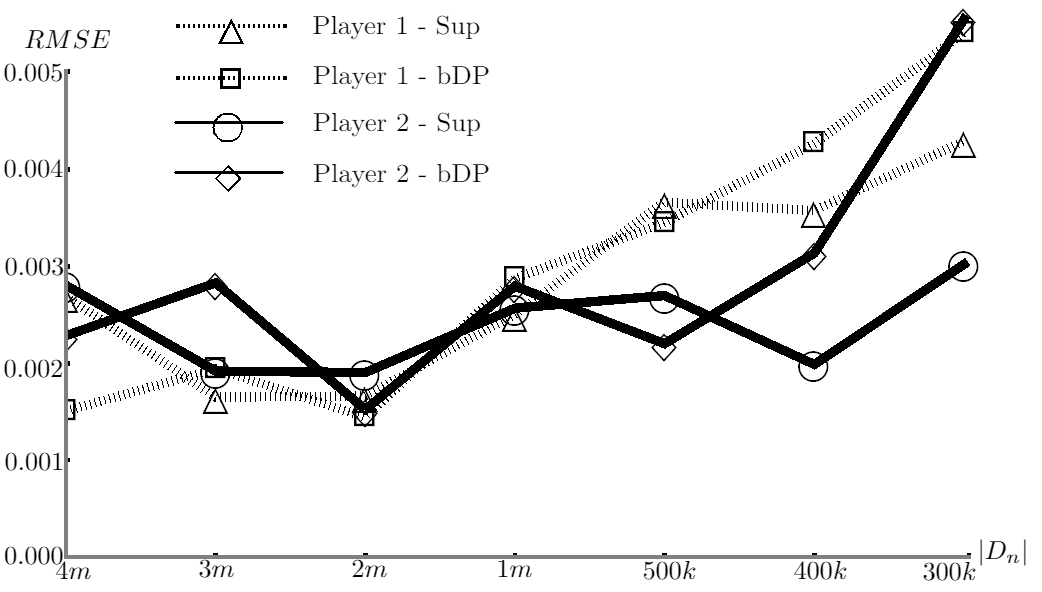}
            \caption{We show the error (RMSE) of self-division (i.e., $\Phi'-\widetilde{\Phi'}$) for both player and privacy methods. }
            \label{fig:rmse}
        \end{figure}
        
        We found, that the RMSE was minimal for both privacy mechanism and player when the players have $|D_n|=\num{2000000}$ ratings. This means, for datasets with density approximately $d=0.05$ $\widetilde{\Phi_n'}$ is the closest to $\Phi_n'$ when $\num{100000}\approx d\cdot|D_n|$ heuristic holds.
        
        %Furthermore, we also found that if the players have more-and-more data then $\Phi_n'-\widetilde{\Phi_n'}$ keeps increasing, while if the players have less-and-less data then this difference keeps decreasing, This effect can be seen in Fig. \ref{fig:sample} where we show $\Phi_n'-\widetilde{\Phi_n'}$ for players with 3/2/1 (lightgray/gray/darkgray) million data.
        
        %This result is in line with our finding in Fig \ref{fig:ratio}: the improvement is less by collaboration when the players have large amount of data, while it is larger when the datasets are small.
        
        \pagebreak
        \section{Playerwise Approximations}
        \label{app:excel}
        
        \begin{table}[h]
            \centering
            \begin{tabular}{|c||c|c|c|c|}
                \hline
                $\widetilde{\Phi_1}$ & $p_2=0.0$ & $p_2=0.2$ & $p_2=0.4$ & $p_2=0.6$\\
                \hline
                \hline
                $p_1=0.0$ & $\num{0.28}$ & $\num{0.26}$ & $\num{0.24}$ & $\num{-0.05}$\\
                \hline
                $p_1=0.2$ & $\num{0.25}$ & $\num{0.16}$ & $\num{0.15}$ & $\num{-0.05}$\\
                \hline
                $p_1=0.4$ & $\num{-0.07}$ & $\num{-0.10}$ & $\num{-0.19}$ & $\num{-0.37}$\\
                \hline
                $p_1=0.6$ & $\num{-1.01}$ & $\num{-1.16}$ & $\num{-1.37}$ & $\num{-1.72}$\\
                \hline
            \end{tabular}
            \begin{tabular}{|c||c|c|c|c|}
                \hline
                $\widetilde{\Phi_2}$ & $p_1=0.0$ & $p_1=0.2$ & $p_1=0.4$ & $p_1=0.6$\\
                \hline
                \hline
                $p_2=0.0$ & $\num{0.17}$ & $\num{0.16}$ & $\num{0.15}$ & $\num{-0.05}$\\
                \hline
                $p_2=0.2$ & $\num{0.14}$ & $\num{0.12}$ & $\num{0.12}$ & $\num{-0.07}$\\
                \hline
                $p_2=0.4$ & $\num{-0.14}$ & $\num{-0.17}$ & $\num{-0.28}$ & $\num{-0.60}$\\
                \hline
                $p_2=0.6$ & $\num{-1.19}$ & $\num{-1.21}$ & $\num{-1.28}$ & $\num{-1.83}$\\
                \hline
            \end{tabular}
            \begin{tabular}{|c||c|c|c|c|}
                \hline
                $\Phi_1$ & $p_2=0.0$ & $p_2=0.2$ & $p_2=0.4$ & $p_2=0.6$\\
                \hline
                \hline
                $p_1=0.0$ & $\num{0.17}$ & $\num{0.14}$ & $\num{0.11}$ & $\num{-0.03}$\\
                \hline
                $p_1=0.2$ & $\num{0.15}$ & $\num{0.12}$ & $\num{0.08}$ & $\num{-0.26}$\\
                \hline
                $p_1=0.4$ & $\num{-0.13}$ & $\num{-0.19}$ & $\num{-0.33}$ & $\num{-0.69}$\\
                \hline
                $p_1=0.6$ & $\num{-1.16}$ & $\num{-1.32}$ & $\num{-1.49}$ & $\num{-2.08}$\\
                \hline
            \end{tabular}
            \begin{tabular}{|c||c|c|c|c|}
                \hline
                $\Phi_2$ & $p_1=0.0$ & $p_1=0.2$ & $p_1=0.4$ & $p_1=0.6$\\
                \hline
                \hline
                $p_2=0.0$ & $\num{0.31}$ & $\num{0.23}$ & $\num{0.17}$ & $\num{-0.05}$\\
                \hline
                $p_2=0.2$ & $\num{0.31}$ & $\num{0.22}$ & $\num{0.11}$ & $\num{-0.18}$\\
                \hline
                $p_2=0.4$ & $\num{-0.14}$ & $\num{-0.16}$ & $\num{-0.22}$ & $\num{-0.52}$\\
                \hline
                $p_2=0.6$ & $\num{-1.13}$ & $\num{-1.25}$ & $\num{-1.30}$ & $\num{-1.85}$\\
                \hline
            \end{tabular}
            \vspace{0.1cm}
            \caption{The approximated privacy-accuracy tradeoff function for both players ($\widetilde{\Phi_1}$, $\widetilde{\Phi_2}$) and its true value ($\Phi_1$ and $\Phi_2$).}
        \end{table}
    \end{appendices}

\end{document}